\documentclass[letterpaper, 10pt, times, preprint]{elsarticle}

\usepackage{hyperref}

\usepackage{mhchem}

\usepackage{lipsum}

\usepackage{tgtermes}
\usepackage{tikz}
\usepackage[]{mhchem}
\usepackage[letterpaper, portrait, margin=0.60in]{geometry}
\usepackage{hhline}
\usepackage{float}
\usepackage{multirow}
\usepackage{graphicx,booktabs}
\usepackage{tabularx}
\usepackage{xfrac}
\usepackage[compact]{titlesec}
\usepackage{wrapfig,booktabs}
\usepackage{mwe}
\usepackage{adjustbox}
\usepackage{caption}
\usepackage{makecell}
\usepackage{rotating}
\usepackage{overpic}
\usepackage{pict2e}
\usepackage{amsmath,array, amssymb}
\usepackage{xcolor,soul}
\usepackage{tcolorbox}
\usepackage{calrsfs}
\usepackage{longtable}
\usepackage{ragged2e}
\usepackage{siunitx}
\usepackage{tikz}
\usepackage{gensymb}
\usepackage{svg}

\hypersetup{colorlinks,breaklinks,
            urlcolor=Maroon,
            linkcolor=Maroon}

\titlespacing*{\subsection}{0pt}{*0}{*0}
\titlespacing*{\subsubsection}{0pt}{*0}{*0}

\begin{document}

\begin{frontmatter}

    
    \title{\large Visualizing High Entropy Alloy Spaces: Methods and Best Practices}

    \author{Brent Vela$^{a}$}
    \author{Trevor Hastings$^{a}$}
    \corref{mycorrespondingauthor}
    \ead{trevorhastings@tamu.edu }
    \author{Raymundo Arróyave$^{a,b}$}
    
    \address{$^a$Materials Science and Engineering Department, Texas A\&M University, College Station, TX, USA 77840}
    \address{$^b$Mechanical Engineering Department, Texas A\&M University, College Station, TX, USA 77840}



\cortext[mycorrespondingauthor]{Corresponding author: Trevor Hastings} 


\begin{abstract}

Multi-Principal Element Alloys (MPEAs) have emerged as an exciting area of research in materials science in the 2020s, owing to the vast potential for discovering alloys with unique and tailored properties enabled by the combinations of elements. However, the chemical complexity of MPEAs poses a significant challenge in visualizing composition-property relationships in high-dimensional design spaces. Without effective visualization techniques, designing chemically complex alloys is practically impossible. In this methods/protocols article, we present a `toolbox' of visualization techniques that allow for meaningful and insightful visualizations of MPEA composition spaces and property spaces. Our contribution to this toolbox are UMAP projections of entire alloy spaces. We deploy this visualization tool-kit on the following MPEA case studies: 1) Visualizing literature reviews, 2) constraint-satisfaction alloy design scheme, 3) Bayesian optimization alloy design campaigns. Furthermore, we show how this method can be applied to any barycentric design space. While there is no one-size-fits-all visualization technique, our toolbox offers a range of methods and best practices that can be tailored to specific MPEA research needs. This article is intended for materials scientists interested in performing research on multi-principal element alloys, chemically complex alloys, or high entropy alloys, and is expected to facilitate the discovery of novel and tailored properties in MPEAs.

\end{abstract}


\end{frontmatter}
\section{Introduction}

Since its advent in 2004 \cite{yeh2004nanostructured} the high entropy alloying paradigm has garnered considerable attention, even being described as reviving metallurgy and alloy design \cite{praveen2018high}. High entropy alloys are composed of 4 or more principal alloy components at concentrations ranging from 5 to 35 at.\% \cite{MIRACLE2017448}. Multi Principal Element Alloys (MPEAs) are an extension of the high entropy alloying paradigm, and refer to compositionally complex alloys without a single principal alloy component but do not necessarily meet any prescriptions for configurational entropy \cite{chen2022focused}. The motivation behind the MPEA-paradigm is to explore the compositionally complex inner regions of alloy spaces. To date, many MPEAs with various attractive properties have been identified due to the vastness and compositional diversity of the MPEA space. Such properties include yield strength \cite{ZHANG2024}, ductility \cite{LI2023144286}, corrosion resistance \cite{NENE2019168}, high/low thermal conductivies \cite{singh2024alloying,bykov2023transport} and coefficients of thermal expansion \cite{bykov2023transport}, and magnetism \cite{VRTNIK2018122}.

While this chemical diversity has allowed the design and discovery of novel alloys, this same chemical complexity makes visualizing composition-property relationships in MPEA systems difficult. The properties of binary alloy systems can be represented on a standard x-y diagram. Making use of barycentric coordinates and the fact that compositional degrees of freedom $n$ is one less than the order of the alloy system $e$, the properties associated with ternary systems can be plotted over a Gibbs-triangle using contour-lines and color maps. Again making use of barycentric coordinates, quaternary systems ($e = 4$) can be represented by a Gibbs-tetrahedron. Regions inside this Gibbs-tetrahedron can be colored or partitioned according to properties within the quaternary system. Such 3D visualizations are difficult to quickly interpret, yet are still possible. However, quinary systems and above $(e \geq 5, n \geq 4)$ cannot be represented in 3 dimensions. Visualizing high-dimensional alloy spaces has been identified as a challenge facing the MPEA community since at least 2017 \cite{MIRACLE2017448}.

Various attempts have been made to visualize high dimensional alloy design spaces. Regarding conventional dimensionality reduction techniques, stacks of 3D psuedo-ternary diagrams can be arranges in a way to show how a varying a 4th compositional dimension affects the remaining 3 dimensions \cite{MIRACLE2017448} however this method is not scalable to arbitrary dimensions. Schlegel diagrams have been suggested as a method to visualize the MPEA space \cite{MIRACLE2017448}; In a Schlegel diagram, a polytope in $d$-dimensional euclidean space ($\mathbb{E}^d$) is represented by a polytope in $\mathbb{E}^{d-1}$. This projected polytope will have polytopal subdivisions (edges and nodes) in the facet. In these diagrams, nodes encode vertices of the polytope while lines encode edges of the polytope. In the case of MPEAs, the composition space can be represented as a $e-1$-dimensional simplex i.e. a generalization of triangles and tetrahedra to higher dimensions. This simplex can be represented in a lower dimension by a Schlegel diagram. However because Schlegel diagrams are only capable of projections from $\mathbb{E}^d$ to $\mathbb{E}^{d-1}$, these diagrams would only be useful for 3-dimensional and 4-dimensinonal composition spaces, i.e. quaternary and quinary systems. Furthermore, with these diagrams the quinary system could only be visualized in 3D space, another limitation.

Graph networks have been used to visualize the coexistence of phases in hyper-dimensional thermodynamic space \cite{MIRACLE2017448,aykol2019network}. In these graph network implementation of phase diagrams, each phase is represented by a node and if two phases coexist at a given T and P their nodes are connected by a line \cite{MIRACLE2017448,aykol2019network}. In a similar vein, via the use of artistic features such as color, line width, and marker shape, so called `Hull Webs' have been used to visualize thermodynamic quantities, i.e. convex hull depth, reaction driving forces, meta-stability, and the likelihood of phase separation \cite{evans2021visualizing}. While this method provides a means for visualizing coexistence of phases and other thermodynamic properties it is not appropriate to visualize arbitrary properties such as price, density etc.

Regarding more sophisticated and interactive visualization techniques, van de Walle et al. \cite{van2022interactive} demonstrated an interactive software capable of visualizing high-dimensional phase-spaces. The authors demonstrated this framework on the 4-dimensional Cantor alloy space. For a given temperature and pressure conditions, this framework begins by randomly sampling a high-dimensional composition space and evaluating the phase equilibria at each sampled MPEA. MPEAs that are determined to consist of a single phase are discarded; these points are discarded as observations of single phase MPEAs do not provide information regarding \emph{phase boundaries}. Next, the MPEAs are grouped based on the phases take part in each equilibrium. Specifically, compositions are grouped based on the end points are the tie-line these MPEAs lay on, and are further grouped based on the phases present at equilibrium. Next, meshed phase-boundary is created. This generates an estimate of the true phase boundary. Once that a high-dimensional phase diagram is generated cross-section of this `high-dimensional' object can be taken. In this way the dimensionality of the phase diagram is reduced. Despite the advantages of this method (accurate representation of high-dimensional phase space), this framework comes at high computational cost. Furthermore this framework is currently limited to visualization phase boundaries and has not be generalize to other alloy properties of interest. While the aforementioned visualization techniques are useful for specific situations, these techniques do not summarize composition-properties in MPEA systems of arbitrary dimensionality.

Of particular interest to this article are the works that used dimensionality reduction techniques such as t-SNE (t-distributed stochastic neighbor embedding) and UMAP (uniform manifold approximation and projection). These techniques aim to project high-dimensional data to a lower-dimensional embedding. Details on these methods are provided in Section \ref{sec:umapMethod}. These methods have been used extensively in alloy design. For example, in their work with generative adversarial networks (GANs), Li et al. \cite{li2024multi} used t-SNE to visualize and compare the high-dimensional data distributions generated by their GANs. t-SNE enabled them to effectively demonstrate how different GAN architectures captured the underlying data distribution of alloy compositions. This visualization technique helped identify areas where the models succeeded or fell short, providing critical insights for refining the generative models to better fit the complex, multidimensional alloy design space. Similarly, in our previous works, we used UMAP to summarize the composition of a chemically diverse data set of additive manufacturing experiments. The result was a diagram that clustered alloys based on their composition, providing a `family portrait' of the database. While the use of t-SNE and UMAP is valid, these dimensionality reduction methods are only trained on a subset of the design space. Consequently, the resulting graphs can be difficult to interpret and often lack the full context of the barycentric nature of alloy design spaces.

In this work we present a modification to conventional UMAP projections. Specifically, we project the entire barycentric design space to 2D. This results in a projection that can be used to visualize chemistry-structure, chemistry-property, and chemistry-performance relationships.  By using UMAP to project an entire barycentric design space, the resultant projection (alloy space UMAP) is much more interpretable than projections that are only aware of subsets of the alloy space.

There is no single visualization technique that is suitable for all scenarios in alloy design. Each visualization technique has their advantages and disadvantages. Thus the design of HEAs requires a toolbox of visualization techniques that allow for the viewing and interpretation of data in high-dimensional composition spaces. The contribution of this work is 2-fold: 1) We formally put forth a visualization technique known as alloy space UMAPs (AS-UMAP). These AS-UMAPs are useful tools that provide viewers with an intuitive overview of chemistry-property and chemistry-property-property relationships in high dimensional barycentric design spaces. 2) We recommend other commonly used visualization techniques and comment on their pros and and cons. These techniques include: compositional box-whisker plots, pairwise plots, chemical signatures / chemical kernel density estimate (KDE) plots, compositional heat-maps, and compositional barcharts.  All of these techniques distill information concerning high-dimensional alloy design spaces to an interpretable figure. We demonstrate these visualization tools on several MPEA design case studies including: 1) visualizing literature reviews, 2) vizualizing databases of alloys, 3) constraint-satisfaction composition-agnostic design of MPEAs \cite{VELA2023118784}, 4) \emph{in silico} Bayesian optimization of yield strength in MPEAs \cite{arroyave2022perspective}. The methods presented in this work are not exhaustive nor applicable to all scenarios that may arise throughout research on MPEAs, but will hopefully be of use to the MPEA community. Finally, we extend this work beyond the MPEA community by demonstrating how AS-UMAPs can be used in any barycentric design space. Specifically, in Section \ref{sec:poly} we demonstrate how UMAP projections of barycentric coordinates can be used in a simple exercise in polymer design.

\section{Methods}
\subsection{Alloy Space UMAP Projections}\label{sec:umapMethod}
In binary alloy systems each composition can be mapped to a single coordinate \{$x_1$\}. In ternary alloy systems each composition can be mapped to a coordinate \{$x_1$, $x_2$\}. Ternary alloy systems and beyond cannot be mapped to 2 coordinates without dimensionality reduction algorithms (DRA). To visualize high dimensional composition spaces, we seek a DRA that can project a set of compositional vectors of size ($e$,1) to coordinate vectors of size (2,1). Recall that $e$ is the order of the alloy system. For example, the chemically complex shape memory alloy (SMA) \ce{Ni_{40}Ti_{20}Pd_{20}Au_{20}} can be represented by the compositional vector \{0.4, 0.2, 0.2, 0.2\}. There is a need to represent this composition (and all compositions in the Ni-Ti-Pd-Au system) with a coordinate pair \{$x_1$, $x_2$\} that can be plotted in 2D i.e. we are reducing the dimensionality of a vector of size (4,1) to a vector of size (2,1). Furthermore, this projection should be reasonably intuitive to interpret.

Different DRAs will use different embeddings to accomplish this goal with specific injective functions (see Ref \cite{wang2021understanding} for more details) such that a given point \{$x_1$, $x_2$, $x_3$, $x_4$\} will be mapped onto a 2D point \{x$_1$,x$_2$\}. The most well-known DRA is a linear mapping technique called Principal Component Analysis (PCA) \cite{greenacre2022principal}. This method is computationally inexpensive as the embedded dimensions are eigenvectors of the covariance matrix of the dataset. PCA can be thought of as a drawing best-fit hyper-ellipsoid to the data. While this DRA can handle very large datasets, it lacks sophistication in its injective function and does not provide meaningful context to how the data points are clustered. Nonlinear DRAs can better retain global and local features of the original dataset, however nonlinear DRAs are computationally more expensive. Two notable DRAs have been developed in recent years that preserve both local and global structure of projected data: T-distributed Stochastic Neighbor Embedding (tSNE) and Uniform Manifold Approximation and Projection (UMAP).

tSNE is an unsupervised machine learning dimensionality reduction technique renowned for its ability to preserve local structures of data. Since Van der Maaten and Hinton introduced tSNE in 2008 \cite{van2008visualizing}, it has been used in various fields such as genetics \cite{li2017application}, astronomy \cite{anders2018dissecting}, and of particular importance to this work, materials science \cite{kirk2022entropy}. tSNE maps data points from high-dimensional space ($x_i$) to lower-dimensional space ($y_i$) while preserving their local structure. This means that if points $x_i$ and $x_j$ are near each other in high-dimensional space, their embeddings $y_i$ and $y_j$ will also be near each other. Conversely, if $x_i$ and $x_j$ are far apart, $y_i$ and $y_j$ will also be far apart. he algorithm achieves this by converting high-dimensional Euclidean distances between pairs of data points into conditional probabilities $P_{j|i}$ using a Gaussian distribution. More details on the algorithm can be found in Van der Maaten and Hinton's original paper. In this work, we use tSNE to embed entire composition spaces into 2D, as shown in Figure \ref{fig:umapgen}.

Similar to tSNE, UMAP (Uniform Manifold Approximation and Projection) is a non-linear dimensionality reduction technique developed by McInnes et al. \cite{mcinnes2018umap}. UMAP is based on concepts from Riemannian geometry. Both UMAP and tSNE create a high-dimensional representation of the data and then optimize a 2-dimensional embedding to be as similar to the high-dimensional representation as possible. While tSNE uses Kullback-Leibler divergence to measure this similarity, UMAP uses a fuzzy simplicial complex. This complex, which can be visualized as a weighted graph where edge weights represent the likelihood of two points being connected, helps learn the manifold structure of the high-dimensional data. UMAP preserves the global structure of the data by approximating this manifold structure and ensures local structure is maintained by connecting each point to at least its first nearest neighbor. More details on the algorithm can be found in Ref \cite{mcinnes2018umap}.

To create an embedding that encompasses the entire alloy space, we use a similar method for both tSNE and UMAP. We grid sample compositions from the MPEA space to generate a hyper-tetrahedron of compositions. This means that the $n$-dimensional MPEA composition space is sampled at uniform increments, including unary, binary, up to and including ($n+1$)-nary. This hyper-tetrahedron of compositions is then processed by the tSNE and/or UMAP algorithm, projecting the high-dimensional composition space to 2D. For a 2D composition space (ternary), an ideal 2D embedding is an equilateral triangle. A 3D composition space (quaternary) should have a square 2D reference embedding, and a 4-dimensional embedding (quinary) should resemble a pentagon, and so forth.

\begin{figure}[htb!]
    \centering
    \includegraphics[width=1\textwidth]{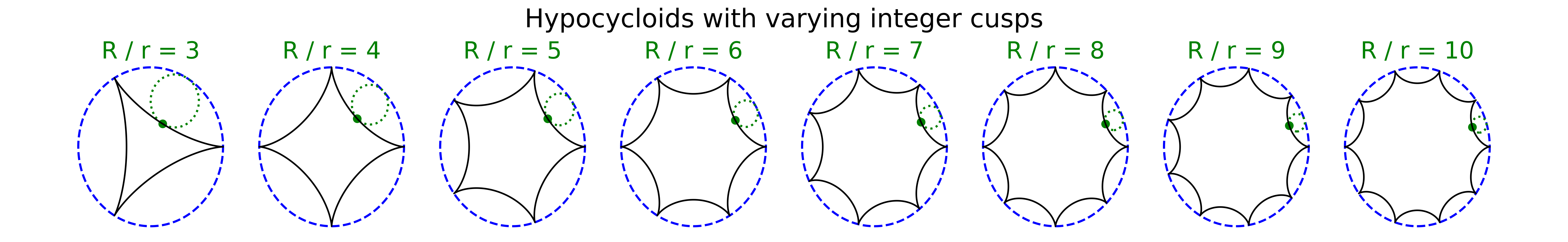}
    \caption{Hypocycloids: The black curve is the shape created when the smaller green circle of radius r rolls around the large blue circle of radius R by tracing the fixed highlighted point.}
    \label{fig:hypocycloid} 
\end{figure}

In the case of UMAP projections, these typically yield hypercycloid counterparts to perfect polygons, as shown in \autoref{fig:hypocycloid}. For example, the UMAP projection of a ternary alloy space results in a deltoid shape, illustrated in the leftmost panel of \autoref{fig:hypocycloid}. This deltoid map can be interpreted similarly to a ternary diagram: each vertex corresponds to a single element, and the middle regions represent compositionally complex alloys comprised of all three elements. High-dimensional hypocycloid projections are interpreted in the same way. For instance, \autoref{fig:umapgen} shows a UMAP projection of a septenary (7-nary) alloy space. Each point represents an alloy with a unique composition. The colored corners in the embeddings indicate compositions with 50\% or more of a particular element, while the central regions contain more `high entropy alloys.'

\begin{figure}[htb!]
    \centering
    \includegraphics[width=1\textwidth]{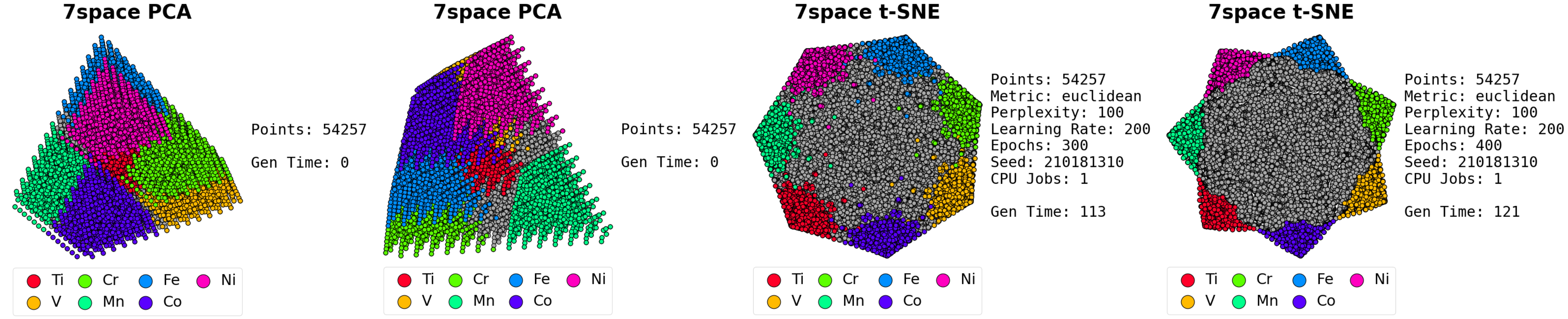}
    \caption{Alternative embeddings: Two PCA examples with different seeds; two tSNE examples with varying perplexity and iterations.}
    \label{fig:pcatsne}
\end{figure}

\begin{figure}[htb!]
    \centering
    \includegraphics[width=0.7\textwidth]{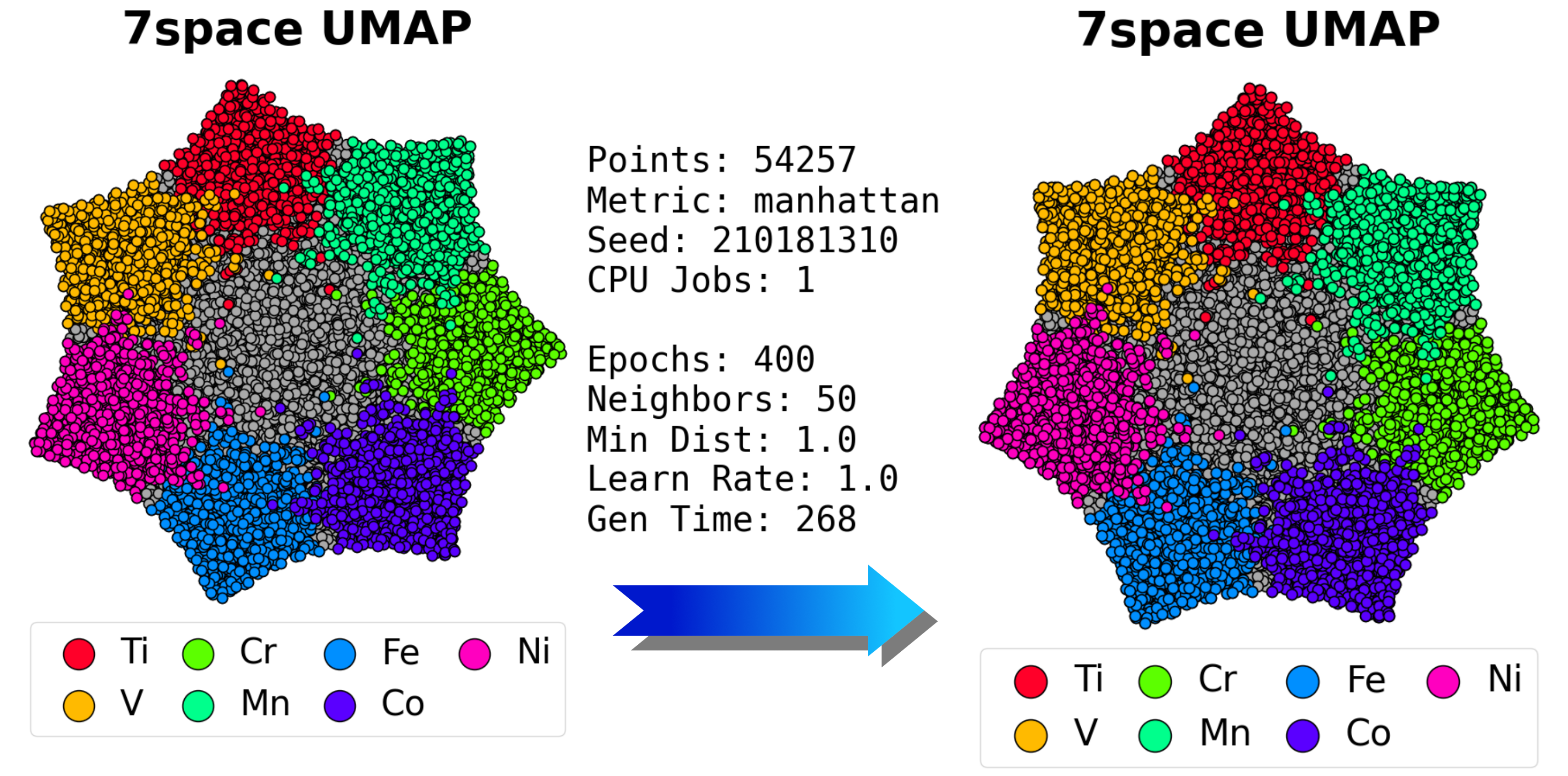}
    \caption{Creation of a UMAP embedding: (left) The raw UMAP output produces coordinates for a local clustering that has an inherently random orientation. Colors represent data points with concentration of that vertex $\geq$ 0.50. It took less than 5 minutes to generate using the listed parameters (spread: default). (right) The embedding after a rotation matrix has been applied to the data, for aesthetic purposes.}
    \label{fig:umapgen} 
\end{figure}

tSNE DRA can also embed a hypercube similarly to UMAP, as shown in \autoref{fig:pcatsne}. In the case of tSNE, the resultant projections also result in polygonal compositional embeddings. Depending on the \texttt{perplexity} and \texttt{n\_iter} parameters chosen (analogous to UMAP's nearest neighbors and number of epochs), it embeds high dimensional composition spaces into regular polygons of $n$ sides or concave polygons with $2n$ sides. These tSNE projections can be interpreted in the same way as UMAPs, i.e. alloys mapped near vertices are rich in a particular element, alloys on the edge connecting two vertices are rich in those 2 elements, and alloys in the central regions are chemically complex.

However, tSNE has a significant shortcoming. tSNE tends to skew compositions with a majority constituent element towards the edges, resulting in overcrowding, as shown in \autoref{fig:pcatsne}. While UMAP also skews compositions with a majority constituent element towards the edges, it better preserves both local and global structures. This allows chemically complex alloys rich in a particular element to be plotted closer to central regions, as shown in \autoref{fig:umapgen}, resulting in less overcrowding. UMAP's resistance to overcrowding is well-documented and is attributed to its superior preservation of structure \cite{njue2020dimensionality}. Consequently, UMAP creates more interpretable alloy space projections than tSNE.

Linear DRA techniques, such as PCA, do not embed composition spaces in an interpretable manner. As shown in \autoref{fig:pcatsne}, while PCA has some clustering ability, it does not intuitively embed compositions like tSNE and UMAP. Specifically, PCA fails to capture global structure in the data. In contrast, UMAP embeds chemically complex alloys in the center of the projection and alloys with a majority element to specific corners of the projection, making it the preferred method for projecting high-dimensional alloy spaces to 2D.

Furthermore, other DRA techniques like kernel PCA (kPCA), and locally linear embedding (LLE), are not suitable for the large numbers of compositions involved in alloy design. UMAP embeddings, on the other hand, do not need a large number of nearest neighbors nor epochs to achieve interpretable embeddings. Using a 7-nary alloy space as an example, \autoref{fig:umapgen} shows the embedding of the Ti-Cr-Fe-Ni-V-Mn-Co alloy space sampled at an atomic resolution of 6.67\%, resulting in a total of 54,257 compositions (excluding unaries). The UMAP in \autoref{fig:umapgen} was generated using a single core on a small laptop (2.8 GHz CPU, 16.0 GB RAM). The embedding was finished in a few minutes. (Note, DRA embeddings have no particular orientation; for aesthetic reasons, we have plotted a rotated version of the UMAP on the right side of \autoref{fig:umapgen}.)

\begin{figure}[H]
    \centering    \includegraphics[width=1\textwidth]{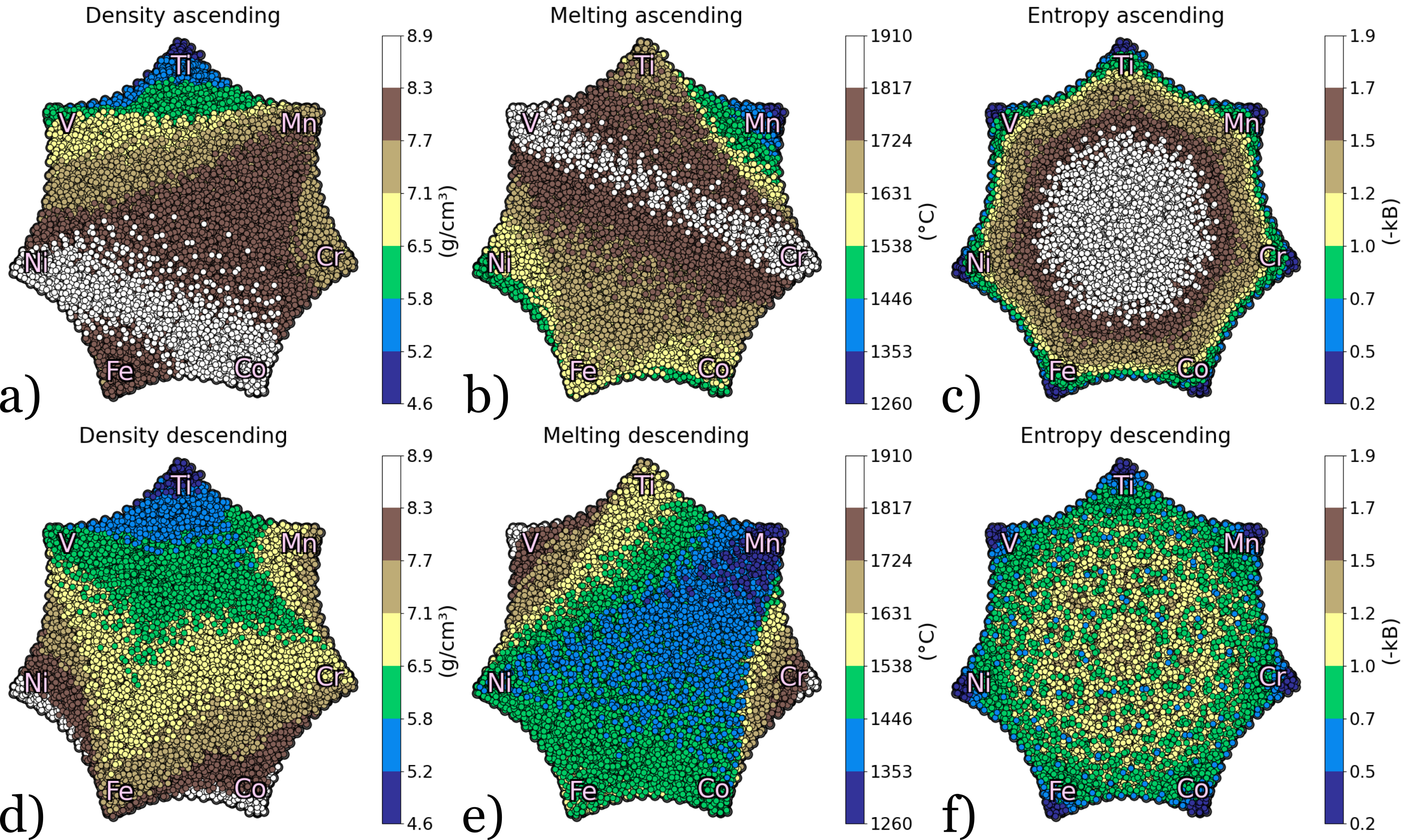}
    \caption{Utilizing a UMAP embedding: Rule of mixtures properties (density, melting point, configurational entropy), plotted in ascending and descending order.}    \label{fig:umapproperty}
\end{figure}

The AS-UMAPs can be interpreted as follows: In  \autoref{fig:umapgen}, each point represents an alloy with a unique composition. The colored corners in the embeddings represent compositions with 50\% or more of a particular element. For example, all alloys colored purple have at least 50 at.\% Co. The central regions represent more `high entropy alloys.' Figure \autoref{fig:umapproperty} shows how properties can be plotted on these UMAPs in order to visualize chemistry-property relationships in chemically complex alloy spaces.

In \autoref{fig:umapproperty}a, the rule-of-mixtures density is plotted as color on the UMAP project. The points are sorted according to ascending density, meaning the densest alloys are plotted on top. The densest alloys are colored white. From \autoref{fig:umapproperty}a it is clear that alloys that are rich in Ni-Co are the densest. This makes sense as the densest elements in the Ti-Cr-Fe-Ni-V-Mn-Co pallet are Ni (8.91g/cc) and Co (8.90 g/cc). \autoref{fig:umapproperty}d shows the same UMAP, however the points are sorted by decreasing density, meaning the least dense alloys are plotted on top. Alloys rich in Ti and V have the lowest densities, as shown in \autoref{fig:umapproperty}d. This makes sense as Ti (4.51 g/cc) and V (6.11 g/cc) have the lowest densities in the elemental pallet.

In \autoref{fig:umapproperty}d, the rule-of-mixtures melting melting temperature is plotted. The alloys with the highest melting points are colored white. In \autoref{fig:umapproperty}b the alloys that have the highest melting temperature are the alloys that are rich in Cr-V binaries. This makes sense as V and Cr have the highest melting points within this elemental pallet (1,910 °C and 1,907 °C, respectively). In the UMAP, these alloys fall near the line from the Cr-vertex to the V-vertex. This white line of alloys has some thickness as the alloys with the highest melt points may also have minor alloying additions of other elements, causing the alloys to be mapped near the Cr-V binary line but not exactly on it. Similarly, \autoref{fig:umapproperty}e shows the same UMAP, however the alloys with the lowest melting points are plotted on top. \autoref{fig:umapproperty}e shows that the elements with the lowest melting points are rich in Mn (1246°C), followed by Ni (1455°C) and Co (1495°C). This is also intuitive as these 3 elements have the lowest melting points in the elemental pallet.


In plot \autoref{fig:umapproperty}c, the ideal configurational entropy is plotted. Alloys with the highest configurational entropy are plotted on top. Alloys with the highest configurational entropies are colored white whereas alloys with the lowest configurational entropies are colored blue. In \autoref{fig:umapproperty}c, compositions with the highest configurational entropies are plotted symmetrically in the center of the UMAP. This is intuitive as elements without a majority element (i.e. compositionally complex alloys) are plotted in the central regions of these UMAPs. These compositionally complex alloys will have a higher configurational entropy by definition.  Likewise, in \autoref{fig:umapproperty}f, it is clear that alloys with low configurational entropy appear near the vertices of the UMAP. This is intuitive as these alloys are rich in a particular element.

With basic knowledge of unary elemental properties, the plot can illustrate overall trends in data as compositions move towards or away from any particular vertex.

The plots in \autoref{fig:umapproperty} can be modified to further slice the dataset. Removing the top 10\% of the data (or 20, 30, etc.) will allow the user to see trends in the middle of the respective legend. Should it be desired, phases can be colored in a similar manner, however this can only be done in the context of the embedding, not topology derived by the laws of thermodynamics. At no point can a DRA be taken as a phase diagram; These UMAP projections are non-unique (random seeds will produce UMAPs with different vertex combinations and slightly different coordinates). This can be advantageous: when the entire barycentric design space is embedded (all the coordinates from 0 to 1), this feature provides flexibility in visualizations. Two vertices may be of higher importance than the rest, such as Ti-Ni in the field of Shape Memory Alloys. If a UMAP generates with these element vertices adjacent, it is likely that their data of interest will be localized to one region of the graph, most of the visual space unused. This is easily rectified as any column can be exchanged and renamed for another, conveniently `rewiring' the UMAP without additional embedding time (again, only possible with all symmetric values from 0 to 1). One example using SMAs in \textit{Results} intentionally separates these two vertices. As a result, creation of UMAPs in this style does \textit{not} require the `hope' that vertices are generated in any particular angular order.

\newpage
\subsection{Compositional Box-Whisker Plots}\label{sec:box_whisker}

\begin{wrapfigure}{R}{0.6\textwidth}
    \centering    \includegraphics[width=0.6\textwidth]{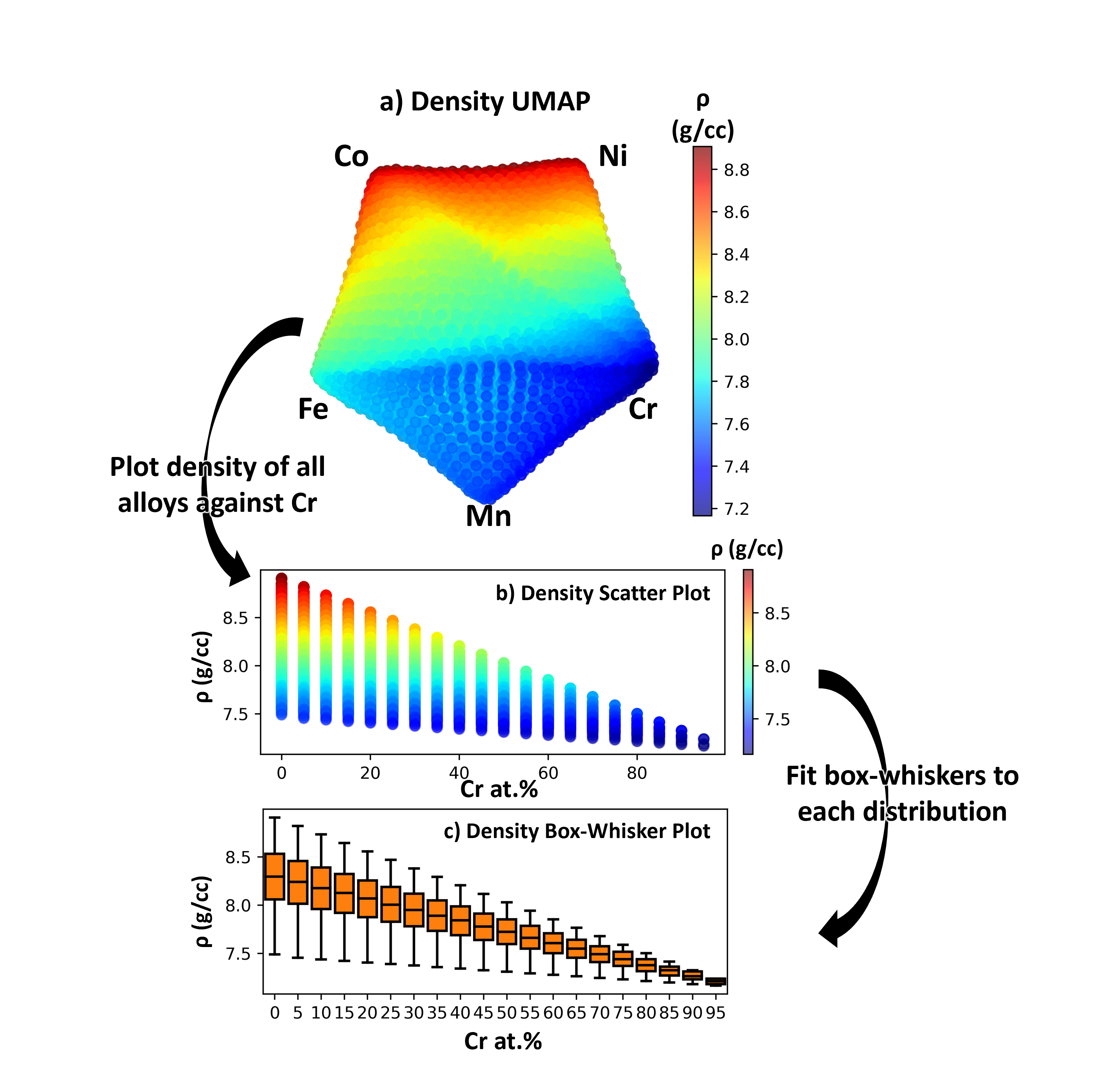}
    \caption{a) UMAP projection of the CoCrFeMnNi alloy space with density plotted on the color axis. b) The density of all alloys in the CoCrFeMnNi alloy space plotted against Cr-content. c) The density distribution of all alloys with a certain Cr-content arranged in order of increasing Cr-content. This plot provides a quantitative summary of the effect of Cr on density.}
    \vspace{-2em}
    \label{fig:demo_boxw}
\end{wrapfigure}

When working in high-dimensional spaces, it is often the case that one wishes to investigate the effect that any alloying component $e$ has on a property. However, this can be difficult in MPEA chemistry spaces because: 1) there are other alloying agents that can confound the effects of the alloying agent of interest and 2) due to the combinatorical vastness of MPEA spaces, 2-D scatterplots can appear overcrowded. Consider the simple example in \autoref{fig:demo_boxw} where the density of the CoCrFeMnNi alloys space is represented with 3 different plots. In \autoref{fig:demo_boxw}a it is evident that alloys that are rich in Cr have the lowest density, however it is difficult to make any quantitative inference about how Cr affects density form this plot. Another way to show the effect of Cr-content on the density of alloys is in \autoref{fig:demo_boxw}b. This plot shows all alloys within the CoCrFeMnNi plotted against Cr-content. From this plot it is evident that Cr addition lowers the density of alloys, and that as Cr content increases, the density of all alloys converges to the density of Cr. However, there many points on the scatterplot that overlap with each other making it it difficult to visualize the distribution of densities at certain Cr-contents. 

Converting the scatter plot to a box-whisker plot allows the summary statistics to be viewed. \autoref{fig:demo_boxw}c shows box-whisker plots as a function of Cr-content. Each box-whisker plot shows the density distribution of all alloys that contain a particular amount of Cr. The first quartile is the bottom portion of the box while the third quartile is the top limit of the box. The interquartile range (IQR) is the length of the box. The ends of the box extend to the maximum and minimum values in the distribution. The diamond-shaped points beyond the whiskers are outliers. With such a plot it is possible to see how measures of center and spread related to a certain property distribution change with composition. This can be achieved in Seaborn using the boxplot function \cite{Waskom2021}. In this way, the effect of alloying agents on properties can be probed quantitatively. The code associated with this toy problem is available at the following repository: \href{https://doi.org/10.24433/CO.7775216.v1}{https://doi.org/10.24433/CO.7775216.v1}.


\subsection{Chemical Signatures}
Often it is desired to briefly summarize the compositions of many alloys without reporting a cumbersome list of compositions. For example, during Batch Bayesian optimization \cite{joy2020batch}, it may be desired to know if candidate alloys converge to a single composition as iterations progress (see Section \ref{sec:BO}). The chemical signature in essence is a histogram that depicts the frequency at which certain elements appear at certain concentrations in a given subset of alloys. For ease of viewing, the underlying histograms are typically omitted and replaced with kernel density estimates (KDEs) that approximate the histograms. These KDEs create unique signature that describe the chemistry of a subset of alloys. These KDE plots can be achieved in Seaborn using the KDE function \cite{Waskom2021}.

\begin{figure}[H]
    \centering    \includegraphics[width=.8\textwidth]{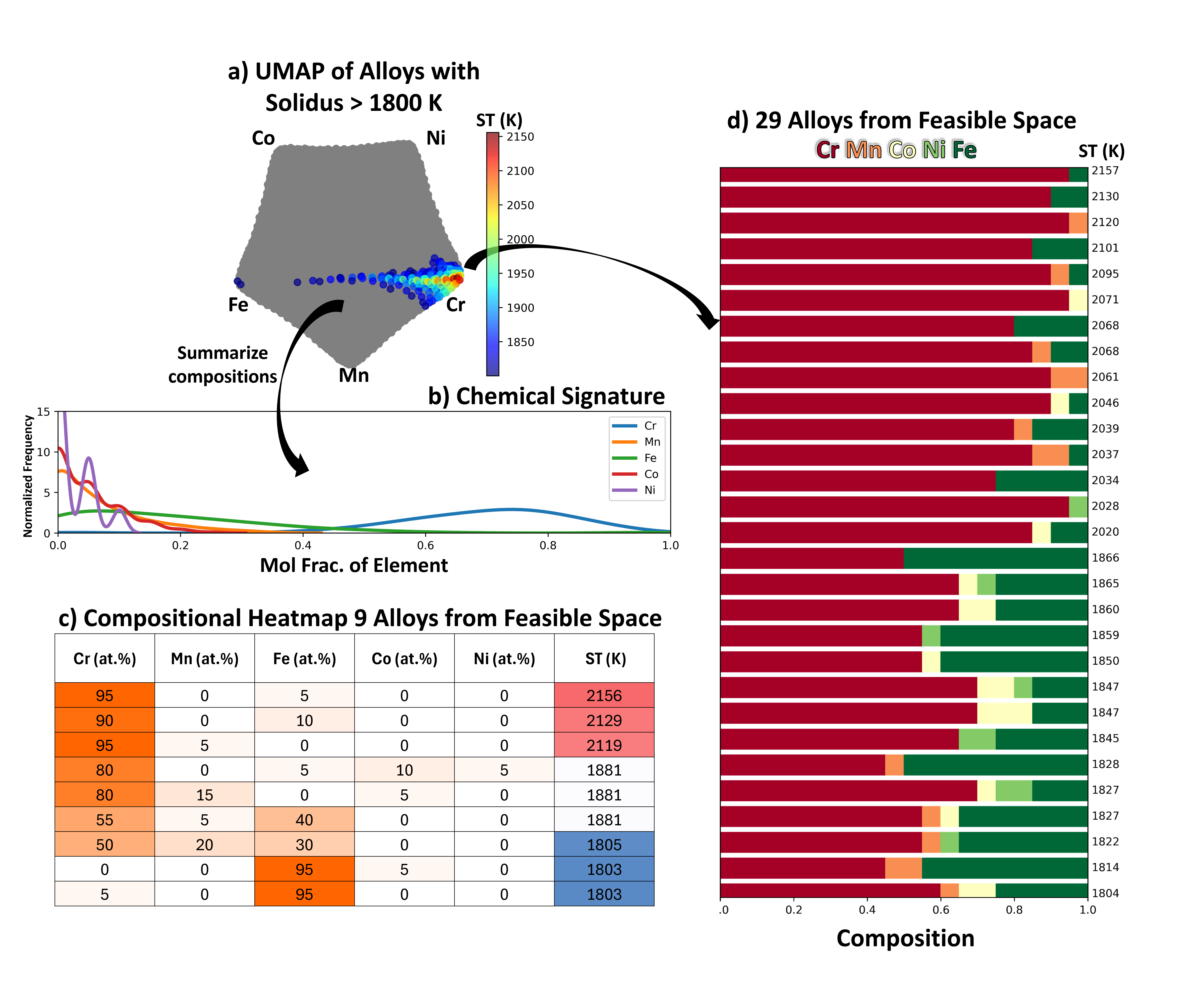}
    \caption{a) UMAP projection of the CoCrFeMnNi alloy space with solidus temperature plotted on the color axis. b) The chemical signature of the 190 alloys that have a solidus temperature great than 1800 K. This plot summarizes these compositions. c) The chemical heatmap of a select few of the 190 alloys that have a solidus temperature greater than 1800K. d) Compositional barchart of a select few of the 190 alloys that have a solidus temperature greater than 1800K.}
    \vspace{-1em}
    \label{fig:demo_chem_sig}
\end{figure}

Consider the subset of alloys within the CoCrFeMnNi alloy space with a solidus temperature greater than 1800K. There are 190 alloys with a solidus greater than 1800 K. These alloys are shown in a UMAP projection in \autoref{fig:demo_chem_sig}a. It is evident from the AS-UMAP that this subset of alloys is rich in Cr and, to a lesser extent, rich in Fe, however, it is hard to make any quantitative inference about the composition of these alloys in this plot. The chemical signature in \autoref{fig:demo_chem_sig}b provides a more quantitative summary of the elemental distributions in the feasible space. The Cr peak near ~75 at.\% indicates that many of the 190 feasible alloys have Cr content near ~75 at.\%. Likewise, the Fe peak near 10 at.\% demonstrates that among the 190 feasible alloys, many alloys contain Fe at ~10 at.\%. Also, note that the Fe signature extends to near 100 at.\%. This indicates that some (but few) alloys are rich in Fe. According to the chemical signatures of the remaining alloying agents, no other element is present in the alloys at concentrations beyond ~45 at.\%.

\subsection{Compositional Heatmaps}
Perhaps the simplest method to visualize composition-property relationships within MPEA spaces is compositional heatmaps. When compositions are presented in tabular formats, it is helpful to add color to each cell to help the viewer recognize a sense of magnitude and scale. This can be achieved with functions as simple as conditional formatting in spreadsheet software such as Microsoft Excel \cite{microsoft2021conditional}. A similar technique is implemented in the visualization software Vital by Kauwe et al. \cite{kauwe2019visualization,belviso2019atomic}. In this visualization, the relative amounts of constituent elements in an alloy space are depicted as color intensity on the cells of a periodic table. An example of a simple compositional heatmap in a spreadsheet can be seen in \autoref{fig:demo_chem_sig}c.

\autoref{fig:demo_chem_sig}c shows the composition and solidus temperature of 9 alloys selected from the 190 alloys with a solidus temperature greater than 1800K. The cells representing the mole percentage of each element are colored according to their value. The highest concentration (95 at.\%) is assigned the darkest orange color while the lowest concentration (0 at. \%) is assigned white. The alloys are sorted according to their solidus temperatures. From \autoref{fig:demo_chem_sig}c it is evident that alloys rich in Cr have melting temperatures well above the 1800K constraint whereas alloys rich in Fe have melting temperatures that are much closer to the 1800K constraint. This makes sense as the melting temperature of Fe is 1811K and the melting temperature of Cr is 2180K. The melting temperature of the other elements is below 1800K, therefore alloys rich in these elements are not represented in the feasible space and thus do not have strong signatures in the compositional heatmap.

\subsection{Compositional Color Barcharts}
A simple way to summarize compositional data and visualize composition-property relationships within MPEA spaces is compositional color barcharts. We took inspiration from Ref \cite{Erps_ink_3dprint} in creating utilizing this method in MPEAs. In compositional color barcharts a colored segment of the bar represents the mole fraction of each element in a particular alloy. Compositional color barcharts are similar to pie charts, showing the relative proportions of various elements within the alloy. However, compositional color barcharts are more interpretable than pie charts as the linear layout of compositional color barcharts allows for straightforward comparison between elements. This linear layout also makes it easy to compare the compositions of a set of alloys. Compositional color barcharts may be stacked and ordered according to a quantity of interest such as MPEA properties. In this way the the effect chemistry-property relationships can be visualized. The code required to make these compositional barcharts are in the following code repository: \href{https://doi.org/10.24433/CO.7775216.v1}{https://doi.org/10.24433/CO.7775216.v1}
.


\autoref{fig:demo_chem_sig}d shows the composition and solidus temperature of a select 50 alloys from the 190 alloys within the CoCrFeMnNi alloy space with a solidus above 1800K. These charts are particularly useful when probing the effect of 2 alloying agents on a property of interest. For better interpretability, the Cr segment is plotted on the far most left and the Fe segment is plotted on the far most right. In this way we see how Fe and Cr increase and decrease as a function of solidus temperature. As the Cr content is depleted and the Fe content is increased the solidus temperature decreases. This can also be seen in the AS-UMAP in \autoref{fig:demo_chem_sig}a where the regions with the highest yield strength (red in \autoref{fig:demo_chem_sig}a) are rich in Cr. As alloys become more chemically complex and rich in Fe the yield strength decreases (blue in \autoref{fig:demo_chem_sig}a). 

\vspace{0\baselineskip}
\subsection{Pairwise Property Plots}\label{sec:pairwise}

\begin{wrapfigure}{R}{0.6\textwidth}
    \centering    \includegraphics[width=0.6\textwidth]{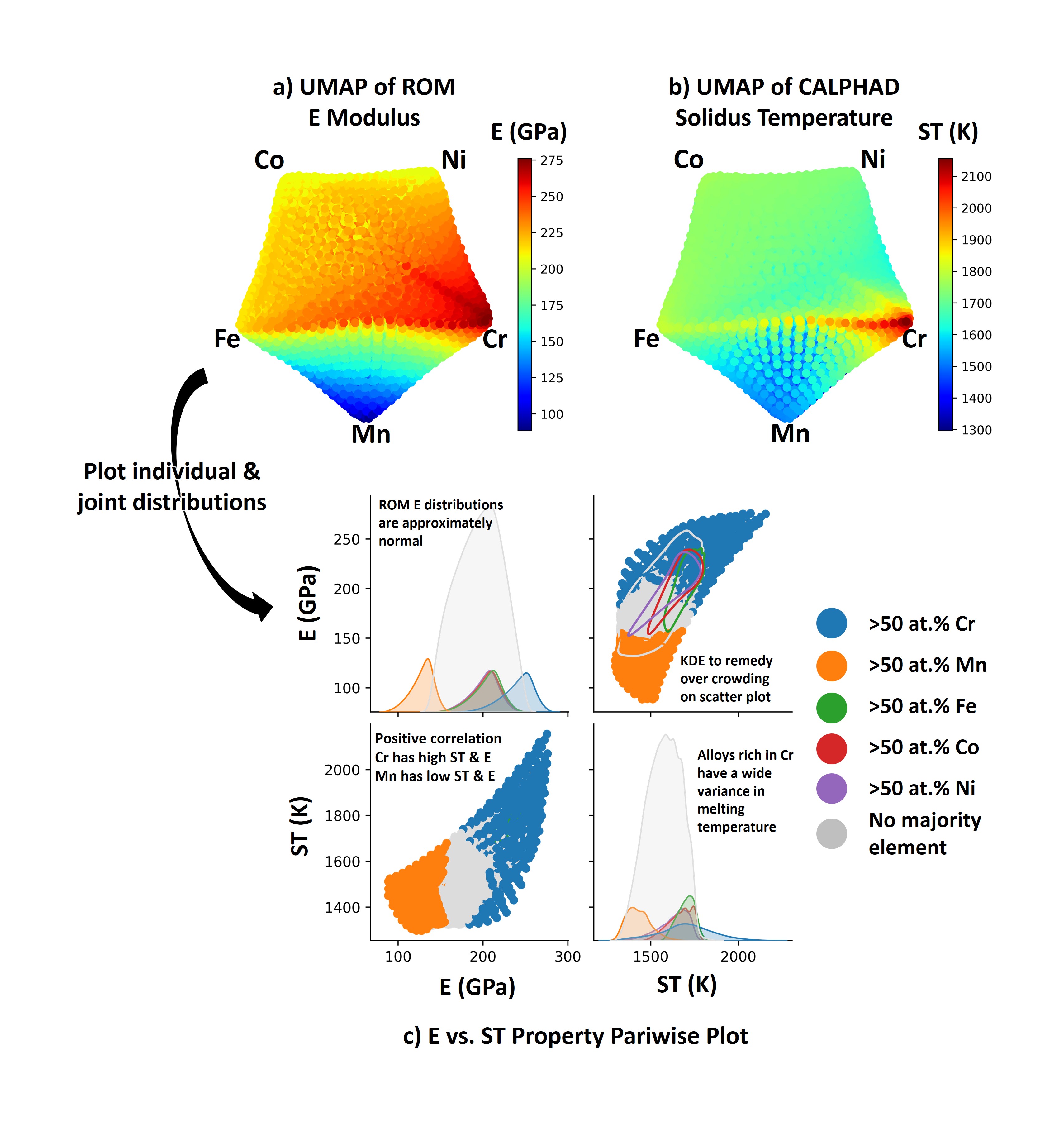}
    \caption{a) UMAP projection of the CoCrFeMnNi alloy space with rule-of-mixtures elastic modulus estimate plotted on the color axis. b) UMAP projection of the CoCrFeMnNi alloy space with CALPHAD solidus temperature estimate plotted on the color axis. c) \st{Pariwise} \textcolor{red}{Pairwise} plot showing the chemistry-property-property relationship between the elastic modulus and the solidus temperature.}
    \vspace{-2em}
    \label{fig:demo_pair}
\end{wrapfigure}

The aforementioned methods are useful for visualizing chemistry-property relationships, however techniques are also needed to visualize property-property relationships in high-dimensional chemical spaces. Pairwise plots consist of a matrix of panels, each combination of which shows a different property-property plot. Pairwise plots have been used extensively in alloy design to show property-property relationships \cite{yang2024convolutional,JAISWAL2021110623,GAO2023105894,khatamsaz2023bayesian,khatamsaz2022multi} however these plots typically do not provide any insight about which compositions have good/bad combinations of properties. To address this we propose modifying pairwise scatter plots by coloring alloys according to their majority element. With this modification, pairwise property plots show both property-property relationships \emph{and} chemistry-property-property relationships. One such pairwise plot is shown in \autoref{fig:demo_pair}.

Consider the scenario in \autoref{fig:demo_pair}a where we have estimated the elastic modulus (E) with the rule-of-mixutres and the solidus temperature (ST) with CALPHAD for every alloy within the CoCrFeMnNi design space. These predictions are plotted on UMAPs in \autoref{fig:demo_pair}b-c to show that we have indeed queried these properties for every alloy in the chemistry space. We can then inspect the E-ST property-property space by creating the pairwise plot in \autoref{fig:demo_pair}c.

In the off-diagonal panels of \autoref{fig:demo_pair}c there is a positive correlation between the predicted E and ST. This correlation between E and ST is known \cite{khakurel2021machine} and thus provides a useful sanity check of this method. Based on the coloring, it can be concluded that Mn has a low combinations of both E and ST whereas Cr has high combinations of E and ST. The scatterplot in the bottom left corner is susceptible to overcrowding. Only 3 alloy classes are visible in the scatterplot (Mn-rich, Cr-rich, and no majority element). To remedy the overcrowding issue, in the top right panel KDE estimates are plotted over the scatter points to better show where alloys rich in a particular element are located in property-property space.

The panels on the diagonals show kernel density estimates (KDE) of the individual property distributions. Alloys with a particular majority element are plotted separately. This is useful when comparing the effect of chemistry on property distributions. For example, in the top left panel the E distribution for Cr-rich alloys has a higher mean than the other distributions. Likewise the Mn-rich distribution has a lower mean than the other elemental distributions. 



\newpage
\section{Results}\label{sec:results}

These techniques can be used to visualize structure-property relationships across high-dimensional spaces in an intuitive way. While the tools in \textit{Methods} are not comprehensive, we believe this suite of visualization techniques is extremely useful when analyzing MPEA design spaces. This section provides a series of case studies utilizing these techniques which showcase a unique material class and property of interest.

\subsection{Utilizing Databases}

\begin{wrapfigure}{R}{0.5\textwidth}
    \centering
    \includegraphics[width=0.5\textwidth]{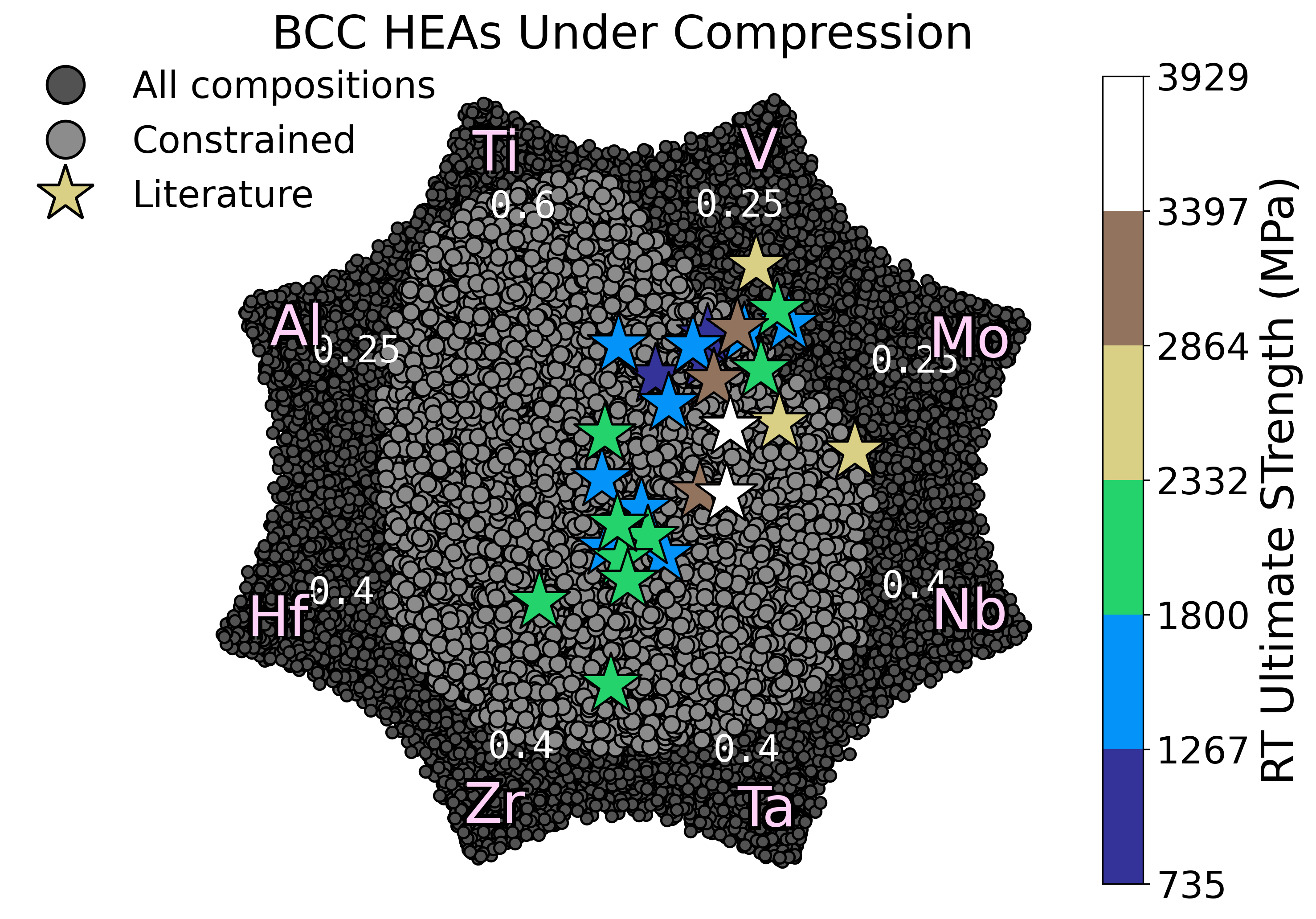}
    \caption{BCC HEAs adapted from the Borg et al. dataset, plotted on a UMAP embedding for the Al-Ti-V-Zr-Nb-Mo-Hf-Ta elemental space. The constrained dataset has maximum atomic fractions according to the annotations near each element vertex. Experimental data is sorted by color according to ultimate strength (MPa) from room temperature compression tests.}
    \vspace{-2em}
    \label{fig:bcchea} 
\end{wrapfigure}

High entropy alloys are perhaps the material class most amenable to UMAP construction as their definition of disorder naturally lends them to a visual capable of integrating several elements. The HEA space is so large that any materials discovery project needs to apply judicious constraints to the design space for practicality. A sensible starting point is to identify a crystal structure of interest, as that will dictate the possible constituents, mechanical procedures like grinding and polishing, and reasonable properties of interest. Unique to HEAs is its definition based on atomic fractions. A HEA design project almost certainly carries atomic fraction maximums at its onset. UMAP embeddings are well suited for overlaying several sets of data points, each of which that are subsets of the previous, based on some project constraint.

Using the database curated by Borg et al. \cite{borg2020expanded}, a selection of BCC HEAs was identified for a UMAP embedding of the 8-element Al-Ti-V-Zr-Nb-Mo-Hf-Ta HEA space. In addition, a list of example elemental maximums was applied to the composition space. For example, candidate alloys in \autoref{fig:bcchea} cannot have Zr present above 40 at.\%. Alloys that pass these compositional constraints are shown in light grey whereas alloys that fail these constraints are shown in dark grey. In \autoref{fig:bcchea}, experimental compressive yield strength data is plotted on top of an AS-UMAP. The color-axis represents the measured yield strength. This case study showcases a single objective problem: the maximization of ultimate strength under compression. With multiple objectives, an array of materials becomes relevant as they dominate all others, the Pareto front of the dataset. Using UMAPs similar to \autoref{fig:bcchea}, one can showcase where the Pareto points are over multiple iterations of experiments. \autoref{fig:bcchea} also provides a visual for the degree to which literature data violates the self-imposed constraints for a project, such as a maximum atomic fraction for an element that would easily oxidize.

In \autoref{fig:bcchea}, a database is used to provide information for an existing project. It provides a literature reference in its chosen objective manifold that the project may or may not intersect. What if we want to inform a project at the design level based on existing databases? This is already a common strategy amongst physics disciplines: cataloging existing experiments across multiple teams to determine which regions can be of scientific interest merely by virtue of being \textit{unexplored}. UMAPs can be an excellent tool for these works - one such example is in the field of shape memory alloys.

Shape memory alloys (SMAs) have seen a large movement towards phase engineering studies, ever since a link was derived between their hysteresis and their crystallographic parameters; by utilizing a formal approach to SMAs with a basis in linear algebra, authors have been able to craft SMAs with abnormally long fatigue lives \cite{RN126}. However, current approaches to find new `near-zero hysteresis' SMAs are dependent on existing literature \cite{TREHERN2022117751}. The large number of compositions and possible tertiary elements can make it difficult to determine which regions of SMA research are lacking in experimental data. Furthermore, a notable roadblock in high-temperature SMA (HTSMA) research has been the lack of experimental data in certain alloy systems, due to the difficult nature of some of the tests. Both of these systems could be supplemented with UMAP visualizations, to gauge a \textit{formal literature assessment} of the field (beyond what is typically limited to long review papers or supplemental tables of compositions).

\begin{wrapfigure}{L}{0.4\textwidth}
    \centering
    \includegraphics[width=0.4\textwidth]{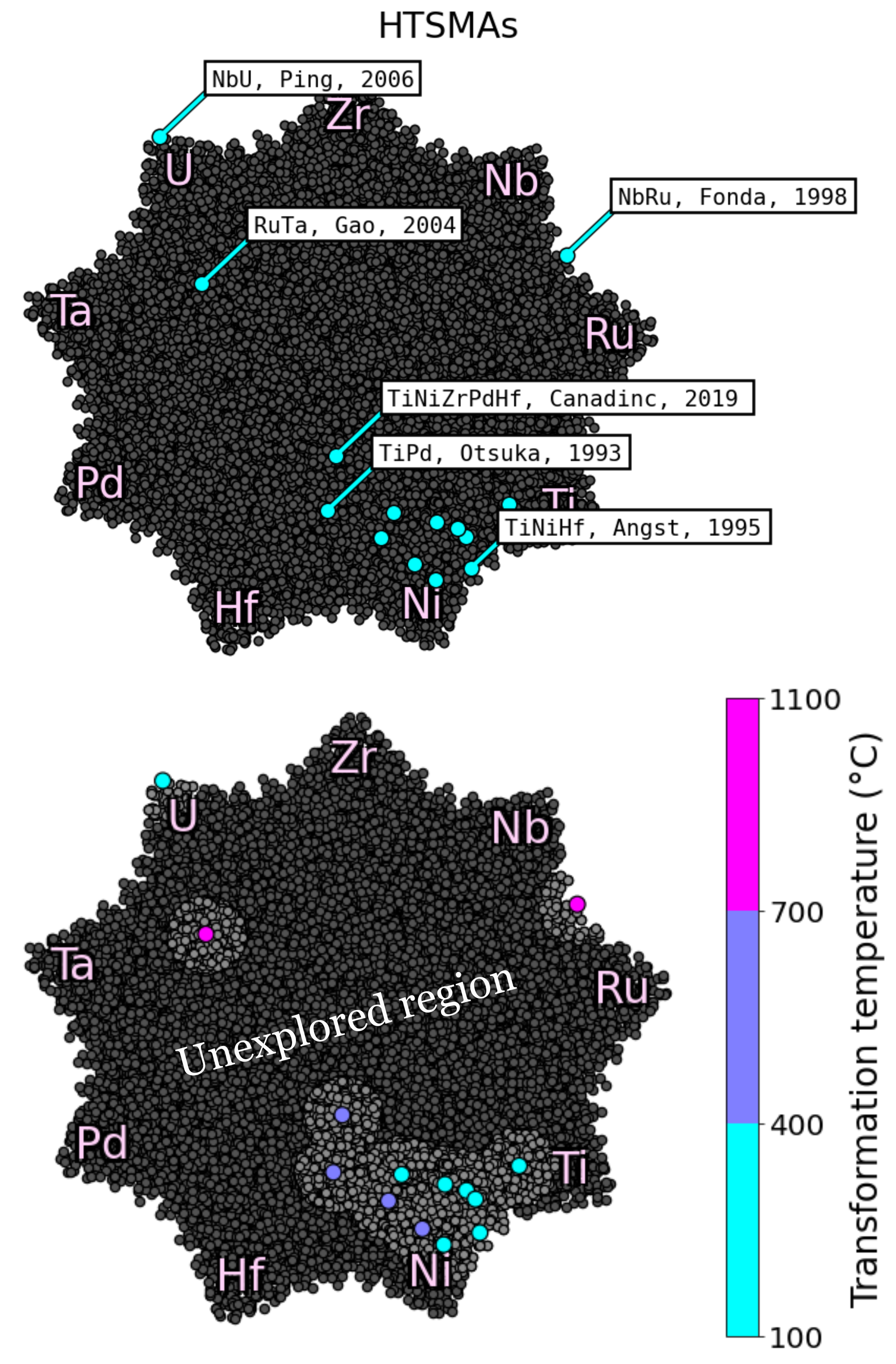}
    \caption{High-temperature shape memory alloys embedded in a UMAP, adapted with data from Karaman et al. and Canadinc et al. (top) The space of Ti-Ni-Zr-Nb-Ru-Pd-Hf-Ta-U HTSMAs with highlighted authors. (bottom) Regions of compositions similar to known alloys in literature are highlighted, and author data points are color-coded according to their three regimes of austenite-martensite transformation temperatures.}
    \vspace{-3em}
    \label{fig:htsma} 
\end{wrapfigure}

In \autoref{fig:htsma}, high-temperature SMAs (HTSMAs) are plotted as an example subset of SMA research. Nine commonly seen elements in HTSMA research were applied to a UMAP embedding based on reported transformation temperatures from a comprehensive HTSMA review \cite{RN122}. \autoref{fig:htsma} (top) includes two databases of reported HTSMAs. A select number of SMAs have been annotated with their elemental subset, a note of the author, and the year of publication. Following the tutorial for UMAPs in \textit{Methods}, a reader can recognize probable regions for compositions on the graph despite not possessing a database of direct compositions. Ti-Ni-X alloys are in a cluster, a binary Ti-Pd example is directly between the Pd and Ti vertices, an HTSMA primarily consisting of uranium is directly by its vertex, and an HTSMA with 5 constituents is closer to the center of the AS-UMAP than any of the other data points.

Note that the positions of components around the vertices are completely arbitrary and \textit{selected by the author}. UMAP generation provides a random solution to the embedding problem built on a specific seed. For typical \textit{asymmetric} embeddings, this arrangement would have to be changed at the embedding level (e.g. an author would have to run the script multiple times with different seeds to obtain one they desire aesthetically). However, in these types of case studies involving atomic fractions, it makes more sense to plot the entire phase space of possibilities (given a maximum atomic resolution). This results in every column of data being \textit{symmetric} i.e. every composition from 0\% to 100\% is somewhere on the graph. After the UMAP embedding has finished, if an author would rather change vertices, they can simply rename and rearrange the columns to their liking. A study specifically focusing on Ni-Ti, for example, might intentionally choose to separate those vertices to better visualize small changes away from a \ce{Ti_{50}Ni_{50}} composition.

In \autoref{fig:htsma} (bottom), the HTSMAs have been categorized by their austenite-martensite transformation temperatures. In addition, data points within the UMAP embedding have been lightly colored within a certain distance of known experimental data, to accentuate compositional regions that are lacking in experiments. Note that this distance is Euclidean \textit{in the UMAP embedding}, using only x and y coordinates and not an Euclidean distance calculated from the dimensionality of the compositions themselves. This visualization has the large advantage of revealing unexplored regions even accounting for additional elements that may not be part of the original dataset.

\autoref{fig:htsma} illustrates several unexplored regions in the SMA design space that authors could pursue. A sufficiently large database of HTSMAs over the past 25 years could be split up into 3 temperature regions across the same UMAP embedding (removing the color bar and using 3 graphs). This would provide a multidimensional view of composition, status in literature, transformation temperature, and a property of choice such as hysteresis width in (°C)---like that shown in \autoref{fig:bcchea}---a very efficient conflagration of information at a glance.

\begin{figure}[b]
    \centering
    \includegraphics[width=0.7\textwidth]{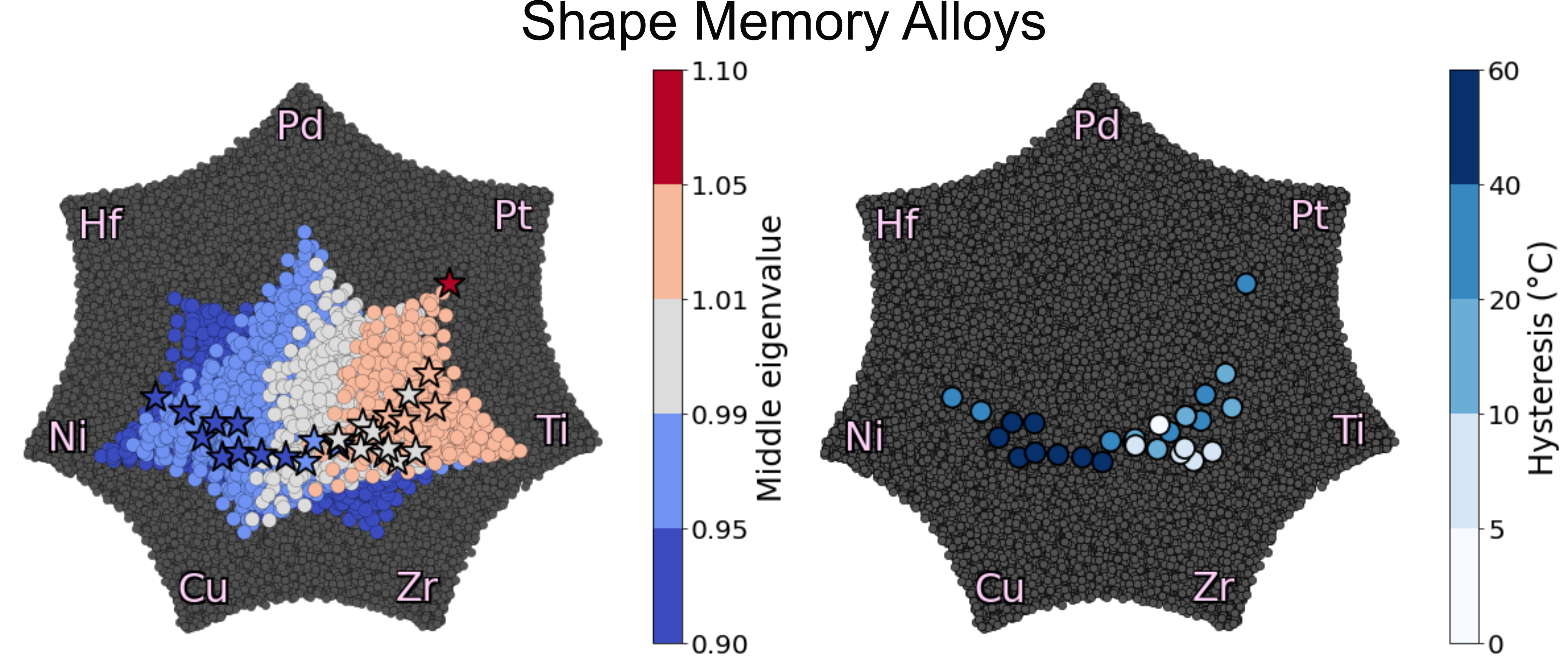}
    \caption{Shape memory alloys embedded in a UMAP including a predictive model for middle transformation eigenvalues, adapted from Zadeh et al. (left) The SMA space with experimental data points overlayed on top of model predictions. (right) Values of Hysteresis widths for those experimentally reported SMAs.}
    \label{fig:sma}
\end{figure}

\autoref{fig:sma} (left) uses a 7-element UMAP representation of NiTi SMAs with several common tertiary additions. In this example the elements' vertices were intentionally separated. Experimental data points (the stars) have been plotted against a background of model approximations of austenite-martensite transformation matrix eigenvalues (the technical indicator for low hysteresis), taken from a database of Shape Memory Alloys \cite{zadeh2024datadriven}. Like before, it is readily apparent which regions of the embedding are lacking in experimental data. In this example, a UMAP provides a novel view of the model's predictions in conjunction with experimental data. Each of the modeled regions in this case overlap, due to the nature of the embedding. \autoref{fig:sma} (right) is a similar UMAP with reported hysteresis values of the same experimental SMAs. The strategy in the UMAPs in \autoref{fig:sma} for SMAs could equally be applied to copper-based SMAs, iron-based SMAs, manganese-nickel-based magnetic SMAs, or simply reduced portions of the extensive NiTi-based SMA literature, focusing on tertiary additions of interest.

\subsection{Constraint-Satisfaction in MPEA Designs Spaces}

Abu-Odeh et al. \cite{ABUODEH201841} showed that the design of high entropy alloys can be framed as a constraint-satisfaction problem.  In constraint-satisfaction design schemes, constraints are applied to an alloy space. The set of alloys that satisfy all constraints is deemed `feasible.' When applying constraints to high-dimensional alloy design spaces it is difficult to visualize which alloys pass/fail certain constraints. In previous work \cite{VELA2023118784} we addressed this visualization challenge by using AS-UMAPs. Specifically, we designed RHEAs for various applications by framing HEA design as a constraint-satisfaction problem. We plot which alloys pass/fail certain constraints on UMAP projections of the design space. In this way, we have a visual summary of the effect of various constraints on the final downselected chemistries. Furthermore we plot the feasible space on UMAPs to show where the `feasible' region lies in the HEA design space. This section will demonstrate how UMAPs (and several other visualization tools) can be used during constraint-satisfaction HEA design schemes.

\begin{wrapfigure}{R}{0.5\textwidth}
    \centering
    \includegraphics[width=0.5\textwidth]{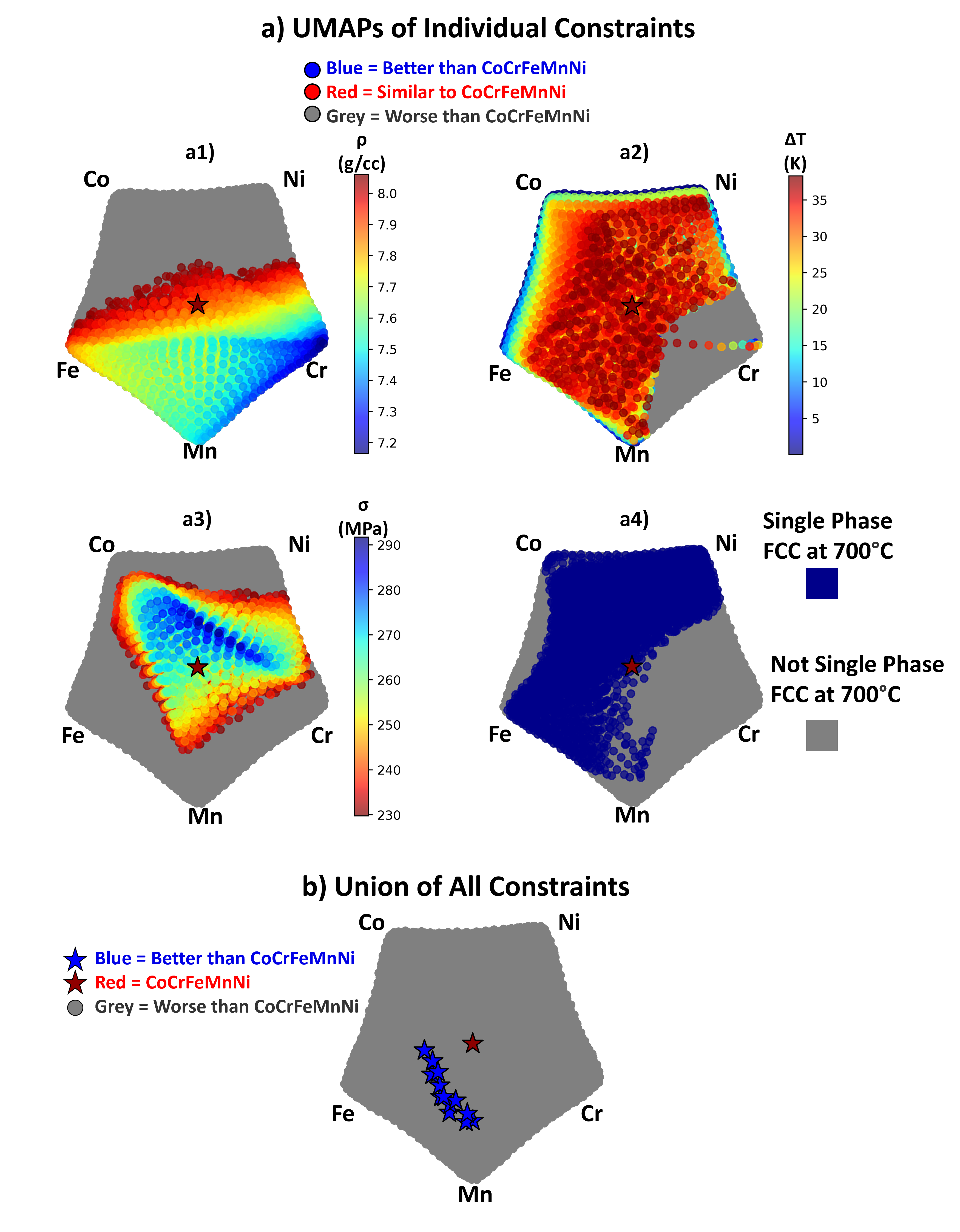}
    \caption{a1) UMAP of the CoCrFeMnNi alloy space that depicts the density constraint. a2) UMAP of the CoCrFeMnNi alloy space that depicts the solidification range constraint. a3) UMAP of the CoCrFeMnNi alloy space that depicts the yield strength constraint. a4) UMAP of the CoCrFeMnNi alloy space that depicts the single-phase FCC at 700\degree C constraint. b) The union of all constraints applied to the CoCrFeMnNi alloy space. The 13 alloys that outperform the equimolar Cantor alloy concerning the 4 aforementioned properties are depicted as blue stars. The equimolar Cantor alloy is depicted as a red star.}
    \vspace{-2em}
    \label{fig:const_sat_UMAP} 
\end{wrapfigure}

Consider a simple \emph{in-silico} constraint-satisfaction design scheme to identify a set of alloys within the Cantor alloy space that has better properties than a benchmark alloy. In this toy problem, the benchmark alloy will be the equimolar Cantor alloy, CoCrFeMnNi. The alloy space is grid sampled at 5 at.\% considering unary to quinary alloys, resulting in 10621 candidate alloys in total. This alloy design scheme aims to identify a set of alloys that have: 1) Single-phase FCC crystal structures at RT for operation at high temperature 2) Low density. 3) Narrow solidification range to avoid processing issues. 4) High yield strength at RT for high-temperature operation. Specifically, feasible alloys must have a predicted single FCC phase fraction of $\geq$ 0.99, a density less than 8.02 g/cc, a solidification range less than 38 K, and an RT yield strength greater than 230 MPa.

The density and solidification range of candidate alloys are predicted using Thermo-Calc's equilibrium CALPHAD simulation \cite{ANDERSSON2002273}. The simulation is conducted using the TCHEA6 database which is appropriate for HEA design spaces, such as the Cantor alloy space. The RT yield strength was predicted using the analytical Varvenne-Curtin model \cite{VARVENNE2016164}. The Varvenne-Curtin model has been widely used by the HEA community to predict the temperature-dependent yield strength of FCC HEAs \cite{SCHNEIDER2020155,MENOU2018185,DEARAUJOSANTANA2022162923,yin2020yield,pei2023toward}. The model is a modification of the theory put forth by Leyson et al. \cite{leyson2016solute}. Specifically, the Varvenne-Curtin model assumes that the rugged energy landscape in HEAs will attract/pin edge-dislocation, hindering their movement through the matrix. The glide of these edge dislocations (and thus softening of the alloy) is facilitated by higher temperatures.

\autoref{fig:const_sat_UMAP} shows the results of this constraint-satisfaction design scheme. The equimolar CoCrFeMnNi alloy (benchmark) is depicted as a dark red star in each UMAP. Its location in the UMAP is intuitive as this equimolar composition lies at the center of the Gibbs hyper-tetrahedron created by this alloy space. \autoref{fig:const_sat_UMAP}a.1 shows the density constraint plotted on a UMAP projection of the CoCrFeMnNi alloy space. Alloys that nearly fail / barely pass the density constraint are colored in red while alloys with low density are colored in blue. In this figure it is clear that Co- and Ni-rich alloys fail this constraint. This makes sense as Co and Ni have the highest densities in the elemental pallet. \autoref{fig:const_sat_UMAP}a.2 shows the solidification range constraint. Cr-rich alloys fail this constraint frequently, as reflected in the AS-UMAP where the Cr-rich region is grey. This makes sense as Cr has a significantly higher melting temperature than the other elements in the pallet. Furthermore, it is evident that compositional complex alloys plotted in the central regions of the UMAP have wider solidification ranges than compositionally simple alloys plotted near the edges and vertices of the UMAP. \autoref{fig:const_sat_UMAP}a.3 shows the RT yield strength constraint. In this AS-UMAP, compositionally complex alloys have a higher predicted yield strength than compositionally simple alloys. This makes sense as the Varvenne-Curtin model is a solid solution strengthening model. Furthermore, alloys rich in Ni and Cr have higher predicted yield strengths. \autoref{fig:const_sat_UMAP}a.4 shows the RT single-phase FCC constraint. Alloys that pass this binary constraint are colored in blue whereas alloys that fail are colored in grey. Alloys rich in Mn and Cr tend to fail this constraint, and this is reflected in Figure 7a.4. This makes sense as Mn and Cr are BCC formers.

\autoref{fig:const_sat_UMAP}b shows the union of these constraints applied to the CoCrFeMnNi design space. When the union of constraints is considered, only 13 alloys are feasible. That is to say, only 13 alloys outperform the equimolar Cantor alloy with respect to the 4 properties of interest. These feasible alloys are compositionally complex and lie in the Fe and Mn-rich region of the design space. In this way, UMAPs can provide a summary of how certain constraints affect the resultant feasible chemistry space. However, UMAPs alone are not sufficient to visualize chemistry-property relationships in HEA design spaces.



Although UMAPs handle overcrowding better than tSNE projections, UMAPs are not completely immune to overcrowding. As a reminder, overcrowding occurs when certain alloys are mapped in such close proximity that they overlap, obscuring other alloys that may have been filtered. Furthermore, UMAPs do not provide a \emph{quantititve} summary of composition-property relationships, but instead only a \emph{qualitative} summary of composition-property relationships. No quantitative relationships about compositions can be made distances between points in such projections. Therefore, relying solely on UMAPs is insufficient for effectively visualizing the correlation between chemistry and properties.

\begin{figure}[H]
    \centering
    \includegraphics[width=1.0\textwidth]{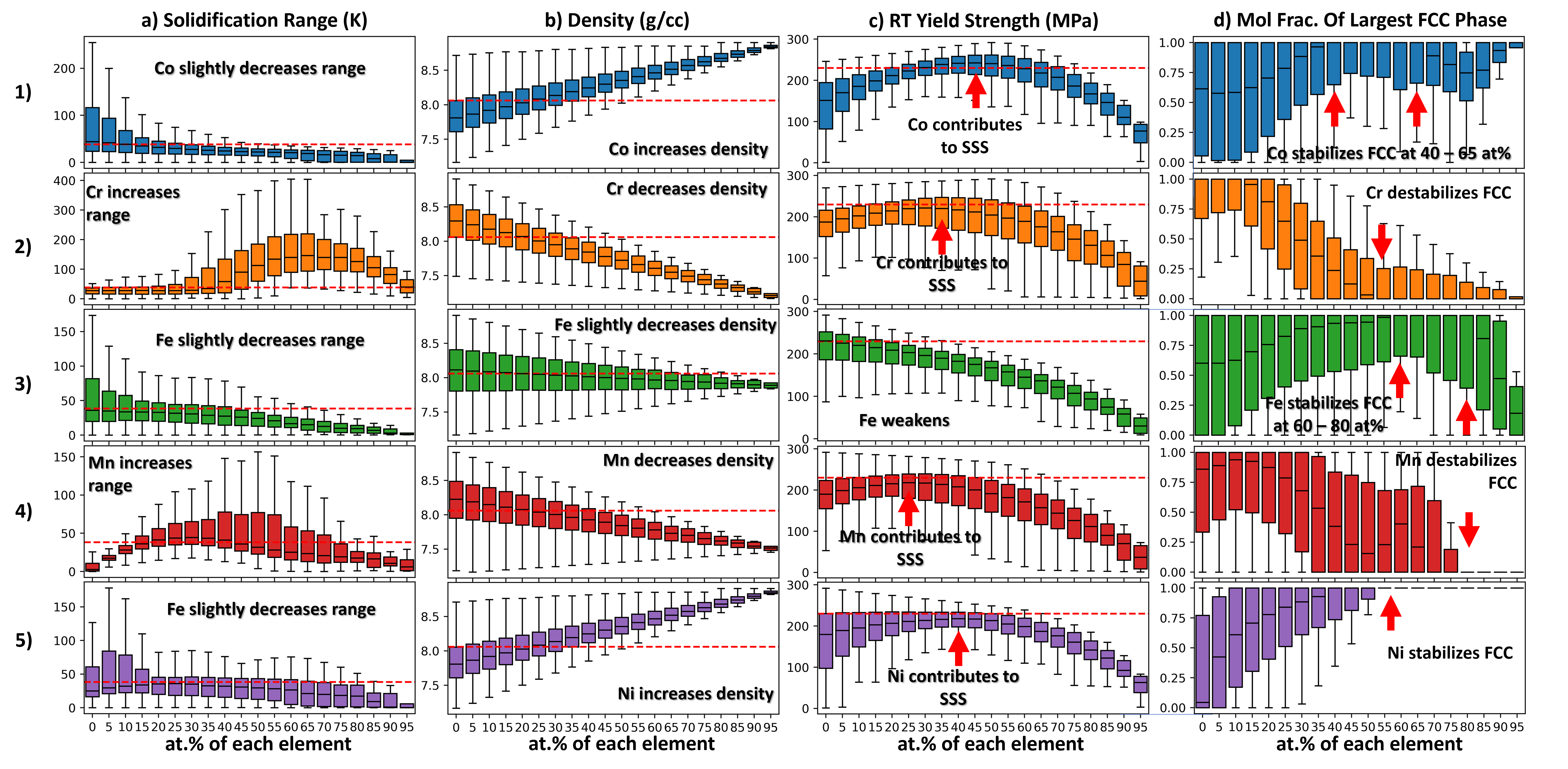}
    \caption{These plots summarize the property distributions using box-whisker plots as a function of individual alloying element concentrations. Each box-whisker plot shows a property distribution when a particular alloying agent is at a certain concentration. Column a shows how the solidification range varies with respect to each alloying agent. Column b shows how the density varies with respect to each alloying agent. Column c shows how the yield strength varies with respect to each alloying agent. Column d shows how the FCC phase stability varies with respect to each alloying agent.}
    \label{fig:design_box_whisker} 
\end{figure}

Another method of visualizing chemistry-property relationships is compositional box-whisker plots (as described in Section \ref{sec:box_whisker}). These plots probe the effect of individual alloying agents on property. The x-axis of each panel in \autoref{fig:design_box_whisker} is the mole fraction of a particular element. When the alloy space is uniformly grid sampled, elements appear at discrete concentration intervals e.g. at 5 at.\% intervals in the case of \autoref{fig:design_box_whisker}. A box-whisker graph is plotted over each interval. These box-whisker plots summarize the property distribution of all alloys that have an element at that specific mole fraction. For example, \autoref{fig:design_box_whisker}1.b shows the effect of varying Co on the density. The box-whisker plot centered over 0 at.\% in \autoref{fig:design_box_whisker}1.b shows the density distribution of all alloys that do not contain Co. Likewise, the box-whisker plot centered over 95 at.\% shows the distribution of all alloys that contain 95 at.\% Co. As chemistry varies along the x-axis the property distribution will vary. In this way we can visually summarize trends between properties and chemistry using simple statistical visualization.

\newpage
In \autoref{fig:design_box_whisker} Column A the solidification range distributions are shown. From Column A it is evident that Co, Fe, and Ni slightly decrease the solidification range of the alloy system. Conversely, Cr and Mn additions increase the solidification range at certain concentrations. However, Cr causes the largest increase in the solidification range by far. This observation is in agreement with \autoref{fig:const_sat_UMAP}a.2 where the Cr-rich region of the UMAP is colored in grey, indicating that that class of alloys frequently fails the solidification range constraint.

In \autoref{fig:design_box_whisker} Column B the density distributions are shown. The trends in this column are linear and easy to interpret as density is known to be accurately predicted using the rule of mixtures. Ni and Co tend to increase the density of Cantor alloys whereas Cr and Mn tend to decrease the density of Cantor alloys. Fe only has a slight effect on density. The IQRs of the density distributions become more narrow as the alloys become richer in a particular element.  The density distributions at 95 at.\% are the most narrow because there are only 4 alloys in each distribution and they are all rich in a particular element and thus have similar densities.

In \autoref{fig:design_box_whisker} Column C the RT yield strength distributions are shown. From Column C it is evident that some elements contribute to solid solution strengthening (e.g. Co, Cr, Mn, Ni) and some elements do not (e.g. Fe). Regarding the elements that do contribute to SSS, these distributions can help us determine the optimal content of each element to achieve SSS. For example, regarding Co, the median yield strength of alloys is maximized when Co content is at ~45 at.\%. Similarly, for Cr this occurs at 35 at.\%. Furthermore, we can see which element has the greatest strengthening effect. From Figure \autoref{fig:design_box_whisker}1.c, it is evident that Co is the most potent strengthener. This is because in the range of 30 to 55 at.\% Co content, the median yield strength is greater than 230 MPa. This is the only element in the design space whose addition causes the median yield strength to exceed 230 MPa over such a wide window of compositions. This is also reflected in \autoref{fig:const_sat_UMAP}a.3 as there are some Co-rich alloys in the feasible region in the AL-UMAP.

In \autoref{fig:design_box_whisker} Column D, the RT single FCC phase fraction distributions are shown. From this figure we see Ni is the most potent FCC stabilizer in the elemental pallet. This is because beyond a Ni content of 55 at.\% all alloys are predicted to have a single FCC phase at RT. Co also promotes a single FCC phase at concentrations between 40 and 65 at.\%. Likewise, Fe promotes a single FCC phase at concentrations between 60 and 80 at.\%. Cr and Mn destabilize the FCC phase. These results are in agreement with the UMAP in \autoref{fig:const_sat_UMAP}a.4.

\begin{figure}[t]
    \centering
    \includegraphics[width=1\textwidth]{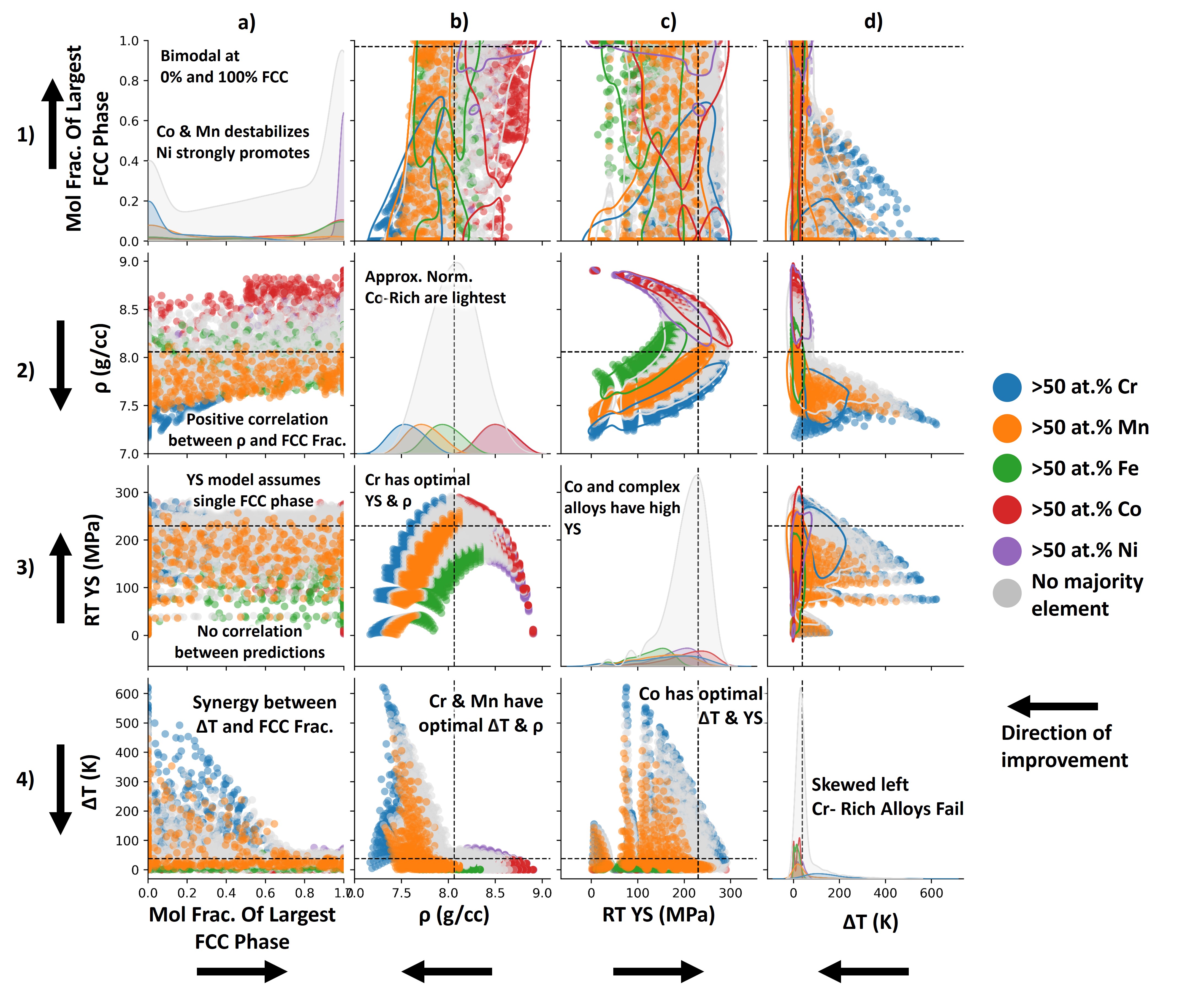}
    \caption{Pairwise property plot showing the chemistry-property-property relationships for this 4-constraint alloy design scheme.}
    \vspace{-1em}
    \label{fig:comp_angost_pairwise} 
\end{figure}

The methods above are useful when visualizing chemistry-property relationships, however it is also important to consider property-property relationships during alloy design. One commonly used method to visualize property-property relationships is pairwise property plots, as described in Section \ref{sec:pairwise}. \autoref{fig:comp_angost_pairwise}  shows the pairwise property plot for the CoCrFeMnNi alloy space. Alloys that have 50 at.\% or more of a particular element are colored according to the legend in the margin of \autoref{fig:comp_angost_pairwise}. The diagonal panels in \autoref{fig:comp_angost_pairwise} depict individual property distributions. The off-diagonal panels depict property-property relationships. Constraints on the properties are depicted with a dashed line.

Regarding individual property distributions, \autoref{fig:comp_angost_pairwise}a.1 shows the mole fractions distributions of the largest FCC phases present in the candidate alloys i.e. if the mole fraction of the largest FCC phase present in a candidate alloy is 100 at.\%, the alloy has a single FCC phase. The distribution in \autoref{fig:comp_angost_pairwise}a is bimodal with peaks at 0 at.\% FCC phase and 100 at.\% FCC phase. The strong peak of alloys that have $>$ 50 at.\% Ni around 100 at.\% FCC phase indicates that Ni-rich alloys are likely to be FCC. This is in agreement with Figure \autoref{fig:const_sat_UMAP}  and \autoref{fig:design_box_whisker} where it was determined that Ni was the most potent FCC promoter in the elemental pallet. Cr (and to a lesser extent Mn) destabilize the FCC phase and thus Cr- and Ni-rich alloys have peaks at 0 at. \% FCC phase.

\autoref{fig:comp_angost_pairwise}b.2 shows the density distributions of candidate alloys. These distributions are all approximately normal. For alloys with a majority element, these density distributions have a mean centered around the density of the pure element. For alloys without a majority element (colored in grey) the density distribution is centered around the density of the equimolar Cantor alloy. The Co-rich density distribution is shifted the farthest to the right indicating that Co-rich alloys are denser whereas the Cr-rich density distribution is shifted the farthest to the left, indicating that Cr-rich alloys are less dense. Few Co-rich alloys pass the density constraint. Alloys on the right side of the Fe-rich distribution fail the constraint. The tail of the Mn distributions fails the constraint. Most of the alloys in the Cr-rich distribution pass the constraint.

\autoref{fig:comp_angost_pairwise}c.3 shows the RT yield strength distributions of candidate alloys. These distributions appear to be left-skewed and log-normal. This constraint filters alloys that have a majority alloying element (e $<$ 50 at.\%). For example, the means of the Ni-, Fe-, Mn-, and Cr-rich yield strength distributions fall below the 230 MPa yield strength constraint. The Co-rich distribution has the most area that falls on the right of the 230 MPa yield strength constraint, indicating that Co-rich alloys have higher yield strengths (according to the Varvenne-Curtin model).

\autoref{fig:comp_angost_pairwise}d.4 shows the solidification range distributions of candidate alloys. These distributions appear to be approximately log-normal. For example, the Mn-rich solidification range distribution appears to be log-normal. Likewise, the no-majority-element solidification range has a log-normal distribution. The distributions of Co, Fe, and Ni, however, have slightly asymmetric tails which might suggest log-normality however these distributions are multi-modal and, therefore cannot be truly log-normal. Cr-rich and no-majority-element alloys fail this constraint frequently. The alloys in the right-side tails of the Mn- and Ni-rich distributions also tend to fail this constraint.

Row 4 shows the relationship between the solidification range and the remaining 3 properties. According to Figure \autoref{fig:comp_angost_pairwise}a.4, There is a synergy between the solidification range and FCC phase fraction in candidate alloys i.e. as the mole fraction of the largest FCC phase increases the solidification range decreases. Regarding the relationship between solidification range and density in Figure \autoref{fig:comp_angost_pairwise}b.4 there is a slight trade-off i.e. as density decreases, the solidification range will tend to increase. Despite this trade-off, Cr- and Mn-rich alloys (and to a lesser extent Fe-rich alloys) have an optimal combination of solidification range and density. Regarding the relationship between solidification range and RT yield strength in \autoref{fig:comp_angost_pairwise}c.4, a trade-off exists i.e. as the yield strength prediction from the Varvenne-Curtin model increases the solidification range will also increase. This is because the Varvenne-Curtin model is a solid solution strengthening model. As the chemical complexity increases the yield strength will increase, but to the detriment of the solidification range.

Row 3 shows the relationship between the RT yield strength and the other properties of interest. There does not appear to be any correlation between the yield strength prediction from the Varvenne-Curtin model and the mole fraction of single FCC phases present in the alloys in \autoref{fig:comp_angost_pairwise}a.3. This lack of correlation may be because the Varvenne-Curtin model is only suitable for single phase FCC solid solutions. The relationship between yield strength and density follows a negative parabolic relationship in \autoref{fig:comp_angost_pairwise}b.3. This parabolic relationship is likely because the Varvenne-Curtin model is a solid solution strengthening model. The yield strength will increase for compositionally complex alloys. These compositionally complex alloys have densities that fall between the densities of their constituent elements, thus the yield strength is maximized when the density is the average density ($\rho$ = 8.02 g/cc). The relationship between yield strength and solidification range is described in the previous paragraph.

Row 2 shows the relationship between the density and the other properties of interest. As shown in Figure \autoref{fig:comp_angost_pairwise}a.2, there exists a slight positive correlation between density and the mole fraction of single FCC phases present in the alloys. The relationships between density and strength and density and solidification range are described in the previous paragraphs.

\begin{figure}[H]
    \centering
    \includegraphics[width=1\textwidth]{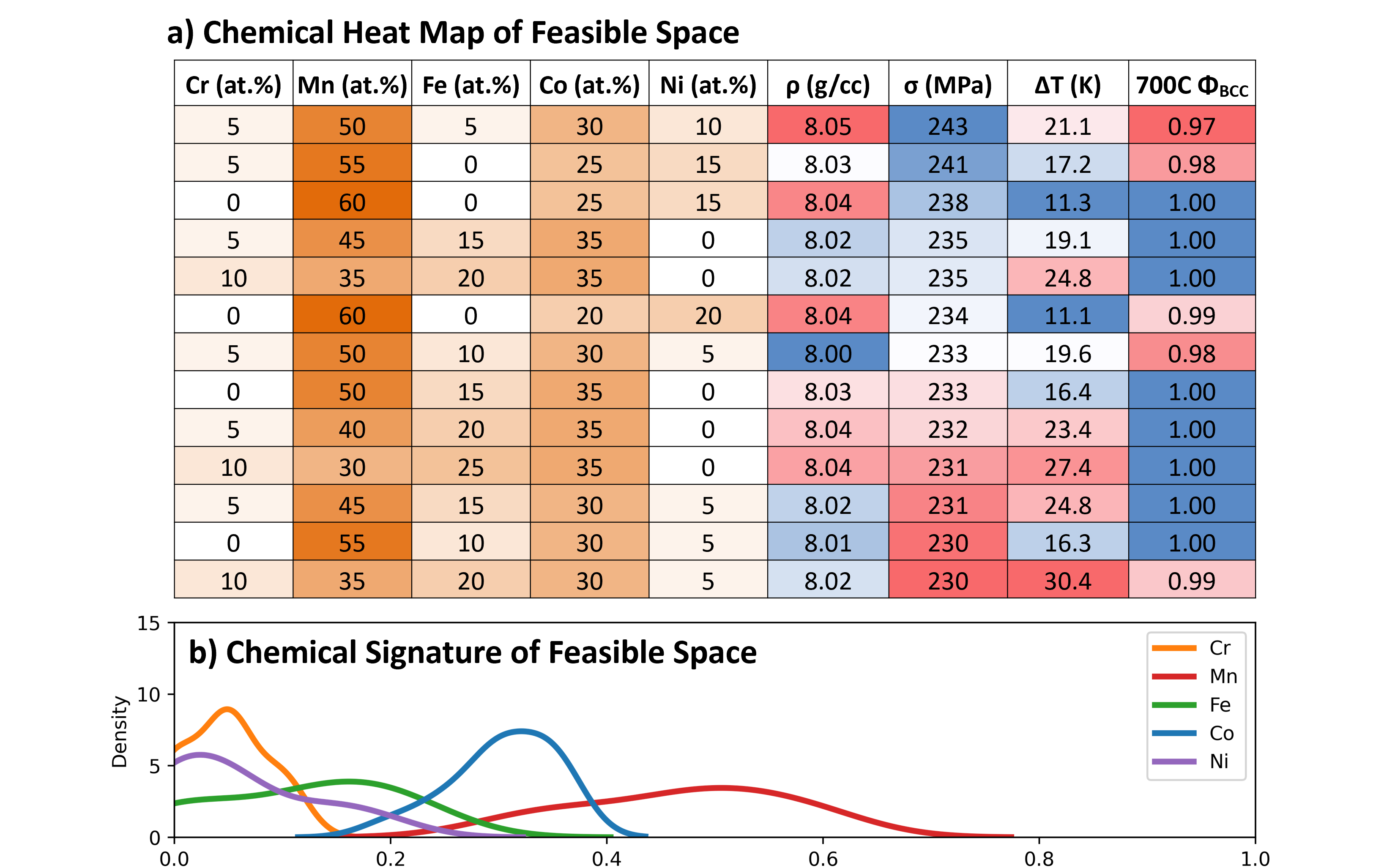}
    \caption{a) Chemical heatmap summarizing the composition and properties of the 13 alloys that outperform the equimolar Cantor alloy with respect to density, yield strength, solidification range, and FCC phase stability. It is evident from the compositional heatmap that the feasible alloys are rich in Mn and Cr to a lesser extent. b) Chemical signature summarizing the composition of the 13 alloys. The Mn signature is shifted to the right indicating that these alloys are rich in Mn. The Mn signature has a large degree of spread, indicating that these 13 alloys have a range of Mn contents. The Co peak is localized around 30 at.\% indicating that all of the feasible alloys have Co contents near 30 at.\%. The other elemental signatures are shifted to the left indicating that these alloys are not rich in these elements. }
    \label{fig:chem} 
\end{figure}

Once the effects of the filters have been probed, the chemistry of the downselected space can be analyzed. Figure \ref{fig:chem} shows different visualizations that summarize the compositions of alloys that pass all the constraints applied in this case study i.e. the set of alloys that outperform the equimolar Cantor alloy with respect to all properties of interest. While 13 alloys is manageable to consider, in many alloy design scenarios the feasible space can be 214 alloys (see Refs. \cite{VELA2023118784}). Therefore techniques that summarize a set of compositions are relevant for alloy design.

\autoref{fig:chem}a is a compositional heatmap. Specifically, the 13 alloys that outperform the cantor alloy with respect to the 4 properties of interest are summarized in tabular form. The cells that contain the composition of each element in the alloy are colored according to their relative amount in the alloy i.e. cells with 60 at.\% are assigned dark orange and cells containing 0 at.\% are colored white. The 4 properties of interest are also tabulated i.e. the density, yield strength, solidification range, and 700 \degree C FCC phase fraction. Each cell in the property column is colored according to its property value. Good values are colored blue and bad values are colored red. For example, in the density column, alloys with the highest density are colored red and alloys with the lowest density are colored blue. In \autoref{fig:chem}a it is evident that the 13 alloys that outperform the equimolar cantor alloy are rich in Mn and to a lesser extent Co. This is in agreement with the AS-UMAP in \autoref{fig:const_sat_UMAP}.

Another method of summarizing the composition of these alloys is the chemical signature shown in \autoref{fig:chem}b. In this figure the frequency at which elements appear at certain concentrations in an alloy is plotted. For example, in this plot we see that if Co appears in the feasible set of alloys, it will appear at concentrations between ~15 at.\% and ~40 at.\%. Likewise it is evident that many of these 13 feasible alloys are rich in Mn. Cr is the least represented element in the feasible space because the Cr KDE is shifted the farthest to the left, toward lower concentrations.

\subsection{Optimization in MPEA Designs Spaces}\label{sec:BO}
Often in optimization problems, the dimensionality of the design space makes visualization difficult. In 1D optimization problems Bayesian optimization can be visualized by plotting the output of a surrogate function (typically a Gaussian Process Regressor). Uncertainties associated with these GPR predictions are typically plotted as shaded regions above and below the prediction from the surrogate model. Typically in the case of GPRs, a $2\sigma$ credible interval is created around the mean prediction from the GPR \cite{garnett_bayesoptbook_2023}. For ternary systems, the surrogate prediction and the uncertainty in the prediction can be plotted on ternary diagrams. Visualization beyond ternary systems becomes cumbersome.  As previously shown, UMAPs offer a method to visualize properties over high dimensional alloy spaces. In the same way, we can visualize the progress of Bayesian optimization schemes in high dimensional alloy spaces using UMAPs. In addition to UMAP projections, in this section we will showcase other visualization techniques that are pertinent to alloy design and Bayesian optimization.



Consider a simple sequential Bayesian optimization scheme with the goal of identifying a set of alloys within the CrNbMoTaVW chemistry space with the highest yield strength as predicted by the Maresca-Curtin model \cite{MARESCA2020235}. The Maresca-Curtin model has been widely used by the MPEA community to predict yield strength. The  Maresca-Curtin model relies on the fact that the random strain fields inherent to MPEAs create a rugged energy landscape that edge dislocations must overcome via thermally activated edge glide. A full derivation of the model is provided in Ref \cite{MARESCA2020235}.

In this optimization scheme, we grid sample the CrNbMoTaVW alloy space at 5 at.\% considering unary to quinary alloys. This sampling results in a grid of 53130 candidate alloys. The goal of the optimization scheme is to locate the alloy with the highest predicted yield strength while minimizing the number of times the Maresca-Curtin model is queried. The GPR surrogate model in this BO scheme is equipped with an additive kernel composed of the anisotropic Radial Basis Function (RBF) kernel and the white noise kernel. The RBF kernel is employed as it is the most common kernel used in GPRs when no prior physics is assumed during modeling. The length scales of the RBF kernel are tuned based on the maximum likelihood as more data is acquired however the length scales are bounded between 2 at.\% and 100 at.\%. The white kernel is added to account for any uncorrelated noise in the data. This kernel is shown in Eqn \ref{eq:kern}. The acquisition function used in the BO scheme is the commonly used expected improvement (EI) metric. This metric quantifies the expected positive difference in yield strength between any candidate alloy (as predicted with the GPR surrogate) and the alloy with the current highest yield strength (as predicted with the Maresca-Curtin model).

\begin{equation}
\label{eq:kern}
k(\mathbf{x},\mathbf{x'}) =  \exp\left(-\frac{(\mathbf{x}-\mathbf{x'})^2}{2l^2}\right ) + \sigma_n^2\delta(\mathbf{x}, \mathbf{x'})
\end{equation}

\begin{figure}[H]
    \centering
    \includegraphics[width=.8\textwidth]{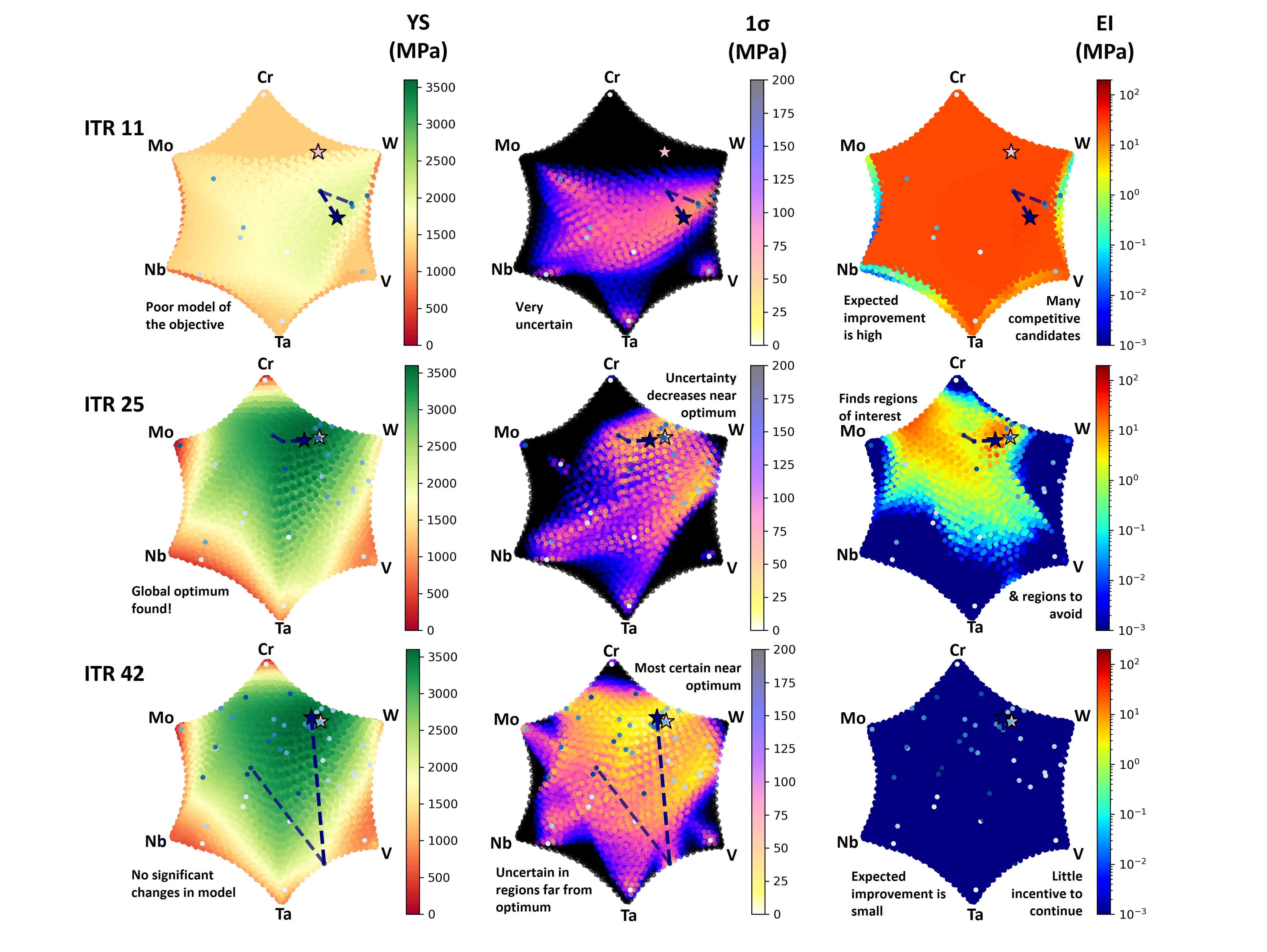}
    \caption{Visualization of \emph{in-silico} Bayesian optimization campaign of the Maresca-Curtin model within the Cr-Nb-Mo-Ta-V-W alloy design space. In ITR 11 the GPR surrogate model is a poor emulator of the ground-truth (the Maresca-Curtin YS model). The uncertainty from the GPR model at ITR 11 is also high, as indicated by the dark coloring on the UMAP. The acquisition function at ITR 11 indicates that many candidate alloys still merit investigation. At ITR 25 the GPR surrogate has improved. Furthermore, the uncertainty in the GPR model has decreased. Likewise, in ITR 25 the acquisition function indicates that Mo-Cr-rich alloys and Cr-W-rich alloys merit investigation. By ITR 42, there is little improvement to the GPR surrogate and the uncertainty has been decreasing over the entire design space. The acquisition function indicates that there are no longer alloys that merit investigation. ITR: Iteration. YS: Yield Strength. 1$\sigma$: One standard deviation. EI: Expected improvement.}
    \vspace{-1em}
    \label{fig:t2} 
\end{figure}

Figure \ref{fig:t2} demonstrates the progression of the BO scheme. The first column of UMAPs represent the objective (yield strength) as predicted using the surrogate function. This represents the current belief about how yield strength varies with chemistry, given the current set of observed data. Green regions represent alloys whose yield strengths are predicted to be higher while red regions represent alloys whose yield strengths are predicted to be lower. In the 11th iteration, the GPR is insufficiently trained and provides a poor approximation of the Maresca-Curtin yield strength. By the 25th iteration the model has improved its model of the Maresca-Curtin yield strength and has found the global optimum (represented by the pink star). The GPR predicts that alloys rich in W and Cr have the highest yield strength. Furthermore, the GPR predicts that pure elements have the lowest yield strength, represented by the red vertices and edges on the UMAP. This is reasonable as the Maresca-Curtin is a solid solution strengthening model. By the 42nd iteration there is little change to the objective model and the BO scheme focuses the majority of its queries on the W- and Cr-rich regions of the alloy space.

The second column represents the uncertainty associated with the prediction from the GPR. Dark regions in the UMAP represent sets of alloys where the GPR is uncertain in its predictions of yield strength. Brighter regions represent sets of alloys where the GPR is less uncertain in its predictions of yield strength. Regions in the alloy space where observations are sparse are thus darker. This is because there is no training data that is compositionally similar to those alloys and the GPR is more uncertain in its predictions. Regions in the alloy space where there are sufficient observations are colored lighter as there is sufficient training data available for these alloys. In the 11th iteration the model is uncertain about its predictions in this design space, and thus the UMAP is colored darker. In the 25th iteration the model is less uncertain about its predictions in the regions near the optimum. This is because, by design, the BO scheme will attempt to focus its queries on the region near the optimum. Fewer queries are made in the V-, Mo-, and Cr-rich regions, indicating that the BO scheme has not sufficiently explored these alloy families. By the 42nd iteration the GPR is more confident in its prediction. Most of the design space has been explored, and the region near the optimum has been exploited.

The third column represents the acquisition function (the EI) at the current iteration. The alloy with the highest EI in the current iteration is then queried at the start of the next iteration. In iteration 11 the EI is high for many alloys within the compositionally complex regions of the design space. The EI is low near the vertices and edges of the UMAP, indicating that the GPR is learning the solid solution strengthening trend in the design space. In the 25th iteration, the EI indicates that the BO scheme is interested in 2 regions in the alloy space. One region is rich in Cr and Mo while the other region is rich in Cr and W. These regions are denoted by bright red colors in the UMAP. It is worth noting that in the 25th iteration, the BO scheme has found the global optimum. Therefore no improvement in the yield strength can be made. However, the BO scheme still `expects' that some alloys have a higher yield strength than the current optimum. Therefore, the optimization scheme will continue querying alloys that are expected to have a higher yield strength than the optimal. By the 42nd iteration, the EI has been decreased significantly. It is evident that there is no incentive to continue the optimization scheme as the expected yield strength improvement for all alloys is on the order of ~1 MPa. These diminishing returns for subsequent experiments indicate the convergence of the BO scheme.

\begin{figure}[H]
    \centering
    \includegraphics[width=1\textwidth]{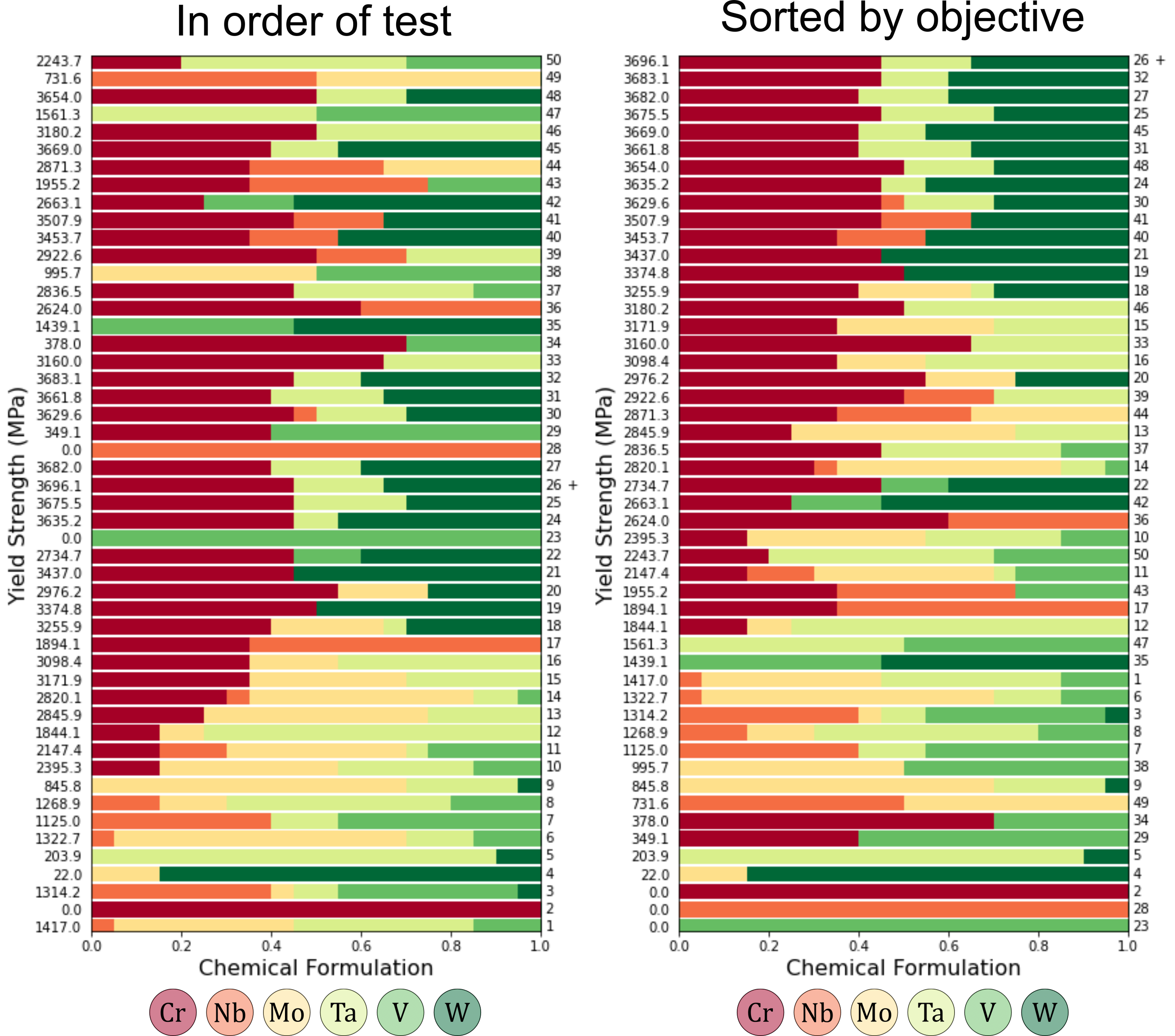}
    \caption{Compositional color bar map of compositions in Figure \ref{fig:t2}, organized by test order and by property order. The maximum is noted with a `+'. It is evident from the left panel that the BO scheme first investigates Cr-rich alloys, then alloys that are rich in Cr and W, and finally begins exploring the space in later iterations. Specifically, the BO scheme investigates alloys that are more rich in Mo. In the right panel where alloys are sorted by objective it is evident that Cr-Ta-W ternaries have the highest yield strength according to the Maresca-Curtin model.}
    \vspace{-1em}
    \label{fig:colorbar} 
\end{figure}

The UMAPs in \autoref{fig:t2} provide an `aerial' perspective of the multidimensional compositional manifold as time progresses, providing the viewer with immediate recognition of trends as optimization progresses. A more direct plot of compositions can be paired alongside these UMAPs to provide quantitative information, without having to resort to a table of numbers that need significant interpretation. In \autoref{fig:colorbar}, the compositions tested in \autoref{fig:t2} are plotted as color bars. This type of plot is particularly advantageous for systems of varying subsystems of elements, as entire degradation mechanisms may differ with the addition or subtraction of a single element. In the right half of \autoref{fig:colorbar}, the tests are sorted by the objective. One can easily see that a particular set of elements, Cr-Ta-W, was more effective than any other combination. The left half of \autoref{fig:colorbar} provides some insight into the candidacy suggestion process of the Bayesian script used. Unary or binary tests 21, 23, 28, 29, 33, 34, 35, etc. show how often the optimization algorithm is willing to `explore' untested regions of the phase space with its given set of hyperparameters. Test 26 reveals the highest objective value ever found; the optimization scheme obviously does not `know' this, and continues to locally test the Cr-Ta-W region. It can be difficult to visualize how far away a composition is from another (in Euclidean distance) when the elements proceed to differ, which is another salient feature of the animation associated with \autoref{fig:t2}, which can be found in the code repository associated with this work: \href{https://doi.org/10.24433/CO.7775216.v1}{https://doi.org/10.24433/CO.7775216.v1}

\subsection{Visualizing Combinatorics: Polymers}\label{sec:poly}

Determining the shape or conformation of polymers has been pivotal to their development for applications; this is typically combined with molecular weight measurements from chromatography and radii of gyration from light scattering experiments \cite{YUAN2015charac} \cite{Terao2004gyration}. As shown in \textit{Methods}, the vertices of a UMAP are inherently unrelated to their materials science concept. They are simply fractions of a whole: this makes the visualization readily accessible to distributions of data, like that of particle sizes, grain sizes, or in this case, molecular weight spreads.

A simple expression for the radius of gyration based on the freely jointed chain model for linear polymers is as follows:

\begin{equation}
\begin{aligned}
<R_{g}> = \sqrt{\frac{1}{6}Nb^{2}}
\end{aligned}
\end{equation}

\begin{figure}[H]
    \centering
    \includegraphics[width=.75\textwidth]{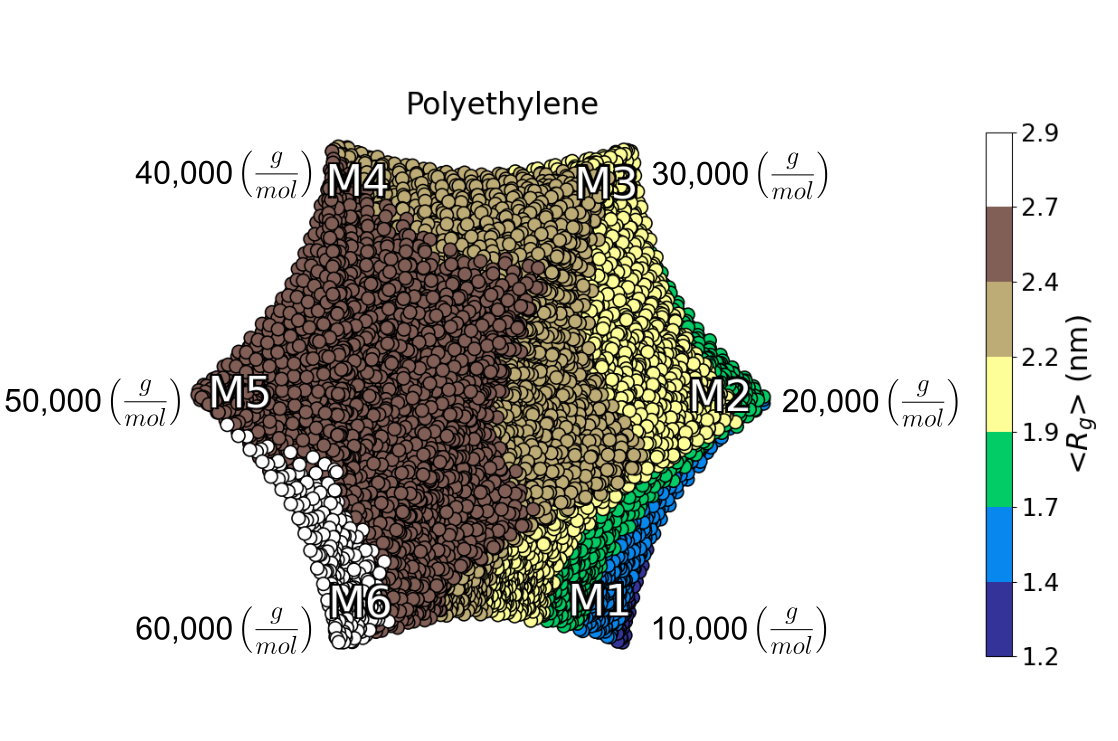}
    \caption{A polyethylene polymer composed of 6 monodisperse fractions, with root mean square radii of gyration calculated across its distribution of molecular weights. The data points at the vertices are equivalent to the calculation of $R|{g}$ for a single chain using that molecular weight.}
    \vspace{-1em}
    \label{fig:polymer} 
\end{figure}

One can then express this as a root mean square of the radius of gyration for a distribution of polymer chains in solution by summing over their fractional occurrences. As an example, fractionated polyethylene, composed of 6 monodisperse fractions, makes up a solution with molecular weights of 10,000 through 60,000 (g/mol). The vertices on a UMAP would then correspond to a completely monodisperse solution, utilizing the equation above, with increasing values of $R_{g}$ with increasing molecular weight. What isn't as obvious at a glance is how this value would change as the distribution does. Using these 6 molecular weights and a resolution of 5\% fractions, \autoref{fig:polymer} shows how the root mean square radius of gyration changes with modifications to the distribution. For the purposes of this model, the persistence length is equal to the length of the monomeric unit (0.154 nm, 28 g/mol).

In cases like these where the visual of a distribution is all that's desired, the ability to precisely label the vertices becomes less pivotal, making this an amenable application for UMAPs of a hypercube with a large number of dimensions, approaching the shape of a circle. Rather than label the monodisperse fractions explicitly, a gradient of molecular weights with radial tick marks could describe many types of polymers. Furthermore, soft matter authors could express properties in terms of other models that better represent polymers with cross-linking or other conformal geometries, via common modifications to the freely jointed chain model \cite{Hamley_soft_matter}. Otherwise identical distributions could be plotted side-by-side, with varying chain models, providing insight into how the choice of model affects predictions of a property over various molecular weight distributions.
\section{Conclusion}\label{sec:conc}

Visualizing high-dimensional composition spaces has been a challenge for the MPEA community. Higher-order MPEA systems cannot be represented on conventional diagrams and require more sophisticated visualization techniques. Some visualization techniques, such as psuedo ternary diagrams are helpful, but cannot probe the effect of individual alloy agents on properties. Other visualization techniques such as Schlegel diagrams and graph networks can be difficult to interpret. Therefore, a suite of intuitive visualization tools are needed for design in compositionally complex alloy spaces. 

In this work, we address this challenge by curating a toolkit of visualization techniques that we have found useful during MPEA design. In this work we present a comprehensive tutorial for this toolkit, detailing the best practices for these visualization techniques. Our unique contribution to this suite of visualization techniques are the UMAP projections of alloy spaces (AS-UMAPs). We provide code demonstrating the utilization of UMAP to visualize high dimensional barycentric design spaces (e.g. alloy spaces). We explain how these AS-UMAPs projections can be used to visualize MPEA composition-property relationships. We believe AS-UMAPs are significant in the context of human-in-the-loop optimization \cite{kusne2020fly} within chemically complex design spaces. Their intuitive nature can enable designers to effectively visualize and navigate complex decision spaces, facilitating more informed and efficient alloy design processes.

In addition to UMAP projections of barycentric design spaces, we demonstrated a suite of other visualization tools that have been used successfully to visualization chemistry-property and property-property relationships in HEA design spaces. We show cased these visualization tools in 5 unique case studies:

\begin{itemize}

    \item  1) We showed the utility of AS-UMAP projections in conducting literature reviews. Specifically, we showed UMAPs can depict explored and unexplored regions in the HEA design space. Furthermore, we showed examples of the chemistry on the experimental yield strength of RHEAs, the beginnings of a formal literature assessment for high temperature shape memory alloys, and model comparisons to shape memory alloy properties.
    \item 2) We showed how AS-UMAPs, compositional box-whiskerplots, pairwise property plots, chemical signatures, and compositional heatmaps can be used to visualize and explain constraint-satisfaction alloy design schemes from start to finish. In this way, chemistry-property, and chemistry-property-property relationships can be visualized. 
    \item 3) We showed how AS-UMAPs and compositional colorbar maps can visualize the progression of iterative Bayesian Optimization schemes. To our knowledge this is the first time a Bayesian Optimization scheme in 5D barycentric design space has been visualized in this manner. We believe UMAP projections of barycentric design spaces can offer useful insights into optimization in high dimensional spaces. The evolution of surrogate model prediction, uncertainity and the acquisiton function can provided designers with information about why the optimization scheme has made certain decisions. This is important for humans-in-the-loop optimization schemes.
    \item 4) Finally we demonstrated that UMAP projections of barycentric design spaces can be applied to other domains in materials science. To this end, we showcase a simple case study using the design of polymers. Specifically, a 5D polymer design space is projected using UMAP.
\end{itemize}

While no single visualization technique is appropriate for all scenarios in alloy design, we believe the visualization tools presented in this work are applicable to \emph{many} scenarios in alloy design and fields beyond metallurgy. We encourage the MPEA community to consider the best and most impactful ways to present their own high-dimensional data.

\section{Data Availability}

The code associated with this work is available at the following repository: \href{https://doi.org/10.24433/CO.7775216.v1}{DOI: 10.24433/CO.7775216.v1}

\section{Declaration of Competing Interest}

The authors declare that they have no known competing financial interests or personal relationships that could have appeared to influence the work reported in this paper.

\section{Acknowledgments}

We acknowledge Grant No. NSF-DGE-1545403 (NSF-NRT: Data-Enabled Discovery and Design of Energy Materials, D$^3$EM) and Grant No. 1746932. We acknowledge support from NSF through Grant No. DMREF-2323611. High-throughput CALPHAD calculations were carried out in part at the Texas A\&M High-Performance Research Computing (HPRC) Facility. We acknowledge support provided by the HTMDEC-BIRDSHOT program (High Throughput Materials Discovery for Extreme Conditions; Batchwise Improvement in Reduced Materials Design Space using a Holistic Optimization Technique) sponsored by the Army Research Laboratory, accomplished under Cooperative Agreement Number
W911NF-22-2-0106.

\bibliographystyle{elsarticle-num}
\bibliography{mybibfile}

\end{document}